\documentclass[aps,prx,preprint,notitlepage,superscriptaddress,floatfix]{revtex4-2}

\usepackage{amsmath,amssymb,braket}
\usepackage{amsfonts}
\usepackage{graphicx}
\usepackage{afterpage}
\usepackage{color}
\usepackage{bm}
\usepackage{tabularx}
\usepackage[normalem]{ulem}

\newcommand{\uf}{u_{(1)}}
\newcommand{\uz}{u_{(0)}}
\newcommand{\dg}{\dot{\gamma}}

\newcommand{\pd}[2]{\frac{\partial #1}{\partial #2}}
\newcommand{\mybra}{\left\langle}
\newcommand{\myket}{\right\rangle}

\newcommand{\auv}{a_{\textrm{uv}}}
\newcommand{\abare}{a_{\textrm{bare}}}
\newcommand{\etaR}{\eta_{\textrm{obs}}}

\usepackage[hypertexnames=false]{hyperref}
\hypersetup{
    colorlinks=true,
    linkcolor=blue,
    citecolor=blue,
    urlcolor=blue,
    breaklinks=true
}

\makeatletter
\let\preprint@abstract@produce\frontmatter@abstract@produce
\def\frontmatter@abstract@produce{%
    \begingroup
    \let\do@output@MVL\@firstofone
    \preprint@abstract@produce
    \endgroup
}
\makeatother

\AtBeginDocument{%
    \setlength{\abovedisplayskip}{9pt plus 2pt minus 3pt}%
    \setlength{\belowdisplayskip}{9pt plus 2pt minus 3pt}%
    \setlength{\abovedisplayshortskip}{5pt plus 2pt}%
    \setlength{\belowdisplayshortskip}{7pt plus 2pt minus 2pt}%
    \setlength{\jot}{5pt}%
}

\begin{document}


\title{Looking at bare transport coefficients in fluctuating hydrodynamics}

\date{\today}
\author{Hiroyoshi Nakano}
\affiliation{Institute for Solid State Physics, University of Tokyo, 5-1-5, Kashiwanoha, Kashiwa 277-8581, Japan}

\author{Yuki Minami}
\affiliation{Faculty of Engineering, Gifu University, Yanagido, Gifu 501-1193, Japan}

\author{Keiji Saito}
\affiliation{Department of Physics, Kyoto University, Kyoto 606-8502, Japan}

\begin{abstract}
Bare transport coefficients in fluctuating hydrodynamics are not directly observable in bulk systems, as hydrodynamic fluctuations inevitably renormalize them into macroscopic values.
In this work, we propose an operational method to determine the bare shear viscosity in two-dimensional dense fluids by focusing on fluid behavior near solid boundaries, where momentum scattering suppresses long-wavelength fluctuations.
Using fluctuating hydrodynamic and molecular dynamics simulations supported by analytical arguments, we show that the viscosity measured near walls directly corresponds to the bare value.
Based on this observation, we construct a practical protocol to extract the bare viscosity from microscopic data and verify its consistency by predicting flow profiles and equilibrium correlations.
We further demonstrate that fluctuating hydrodynamics quantitatively reproduces fluid behavior down to atomic length scales.
\end{abstract}

\maketitle
\clearpage

\section{Introduction}
\label{sec1}

\afterpage{%
\begin{figure}[t]
\begin{center}
\includegraphics[scale=1.00]{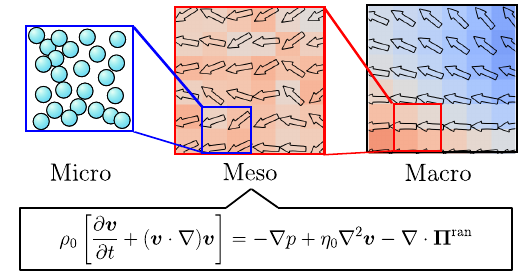}
\end{center}
\vspace{-0.6cm}
\caption{
Schematics of fluid dynamics across scales: microscopic (molecular), mesoscopic (with fluctuations), and macroscopic (Navier–Stokes).
Arrows indicate flow velocity (red: fast, blue: slow).
At the mesoscopic scale, thermal fluctuations require a fluctuating hydrodynamic description.
Symbols denote the density $\rho_0$, velocity $\bm{v}(\bm{r},t)$, pressure $p(\bm{r},t)$, bare viscosity $\eta_0$, and stochastic stress $\bm{\Pi}^{\mathrm{ran}}(\bm{r},t)$, which represents thermal fluctuations.
}
\vspace{-0.5cm}
\label{fig0}
\end{figure}
}

Predicting the collective behavior of fluids requires an appropriate level of description.
While the deterministic Navier–Stokes equation describes macroscopic dynamics, mesoscopic behavior is influenced by thermal fluctuations originating from microscopic degrees of freedom.
Fluctuating hydrodynamics extends the classical framework by incorporating stochastic noise~\cite{Landau1959-eu} and provides a description of such mesoscopic regimes~\cite{Onuki1979-jt, Hohenberg1992-lr, Onuki2002-ff, Bell2010-ly, Balakrishnan2014-vg, Spohn2014-dw, Spohn2016-by, Kim2017-cg, Narayanan2018-fr, Barker2023-ua, Srivastava2023-nx, Tauber2002-nb, Nakano2019-ww, Nakano2020-cu, Kado2024-no, Sasa2024-ln}.
This framework has been used to study long-range correlations in nonequilibrium steady states~\cite{Ronis1982-uk, Lutsko1985-zb, Garrido1990-dy, Dorfman1994-cl, De_Zarate2006-xw, Bedeaux2015-lu, Sengers2024-gz} and the influence of thermal fluctuations on turbulent flows~\cite{Bell2022-en, McMullen2022-xb, Bell2022-jo, Bandak2022-rk, Qin2024-gl, Bandak2024-al, Ishan2025-fw}.
It also provides a natural setting for studying nonlinear fluctuation effects across scales~\cite{Mazenko2006-jq, Das2011-ao, Yakhot1986-mp, Kardar1986-td, Frey1994-wl}, with applications to dynamical critical phenomena and anomalous transport in low-dimensional systems~\cite{Hohenberg1977-zt, Onuki2002-ff, Forster1976-ea, Forster1977-lr, Narayan2002-zo}.

In fluctuating hydrodynamics, thermal fluctuations are incorporated by adding stochastic noise to the Navier–Stokes equation (see Fig.~\ref{fig0}), which introduces bare transport coefficients ($\eta_0$)~\cite{Zwanzig1961-co, Zubarev1983-vd, Mori1973-xn, Kawasaki1973-ql, Fujisaka1976-ir, Espanol2015-ut, Saito2021-rw}.
These bare coefficients are the parameters appearing in the stochastic hydrodynamic equations before the effects of hydrodynamic fluctuations are absorbed into macroscopic transport coefficients.
They therefore generally differ from the macroscopic transport coefficients in deterministic hydrodynamics.

However, bare transport coefficients are difficult to determine~\cite{Donev2011-hf, Donev2011-dv, Espanol2015-ut, Peraud2017-xt}.
In bulk fluids, mode-coupling theory (MCT) and renormalization group (RG) analyses show that transport coefficients are inevitably dressed by fluctuations, so that measurements probe only renormalized values.
In this sense, bare transport coefficients are not directly observable in bulk systems.
Although projection operator methods provide formal expressions for bare coefficients~\cite{Zwanzig1961-co, Zubarev1983-vd, Mori1973-xn, Kawasaki1973-ql, Fujisaka1976-ir, Saito2021-rw}, they are not practical for quantitative evaluation.
These issues motivate the need for an operational approach to determine bare transport coefficients.

In this paper, we propose an operational method to determine the bare shear viscosity in two-dimensional (2D) dense fluids.
We use molecular dynamics simulations of particles confined to two dimensions as a representative system.
Determining the bare viscosity from microscopic dynamics has remained challenging, particularly in 2D where macroscopic transport coefficients exhibit system-size-dependent behavior.
Because our proposed method is operational, it can in principle be applied beyond numerical simulations.
It is therefore relevant to experimental realizations of 2D fluids, including soap films~\cite{Brogioli2017-hf} and electron fluids in ultra-clean graphene~\cite{Gooth2018-bs, Sulpizio2019-bq, Polini2020-gr, Fritz2024-yb}.

Our key idea is to focus on fluid behavior near solid boundaries rather than in the bulk.
We show, using fluctuating hydrodynamic simulations and perturbative analysis, that solid walls suppress the long-wavelength fluctuations responsible for renormalization.
As a result, the local fluid dynamics near the wall is governed by the bare viscosity.
This suppression can be understood as a consequence of momentum scattering at the boundary, which breaks local momentum conservation and inhibits the development of long-range velocity fluctuations.

Based on this observation, we propose an operational protocol to determine the bare viscosity directly from microscopic dynamics.
By fitting flow profiles observed in MD simulations to fluctuating hydrodynamics, the bare viscosity can be extracted.
The resulting value consistently describes independent flow configurations, confirming its physical validity.

In addition to determining the bare viscosity, we examine the role of the spatial discretization scale, hereafter referred to as the ultraviolet (UV) cutoff length, which acts as a parameter in fluctuating hydrodynamics.
Our results show that this parameter affects the quantitative description of transport, reflecting the scale at which the coarse-grained hydrodynamic description is defined.
We find that fluctuating hydrodynamics quantitatively reproduces fluid behavior down to length scales comparable to the atomic diameter.
This supports the interpretation of the bare viscosity as a well-defined parameter associated with the smallest physically relevant scale.

The paper is organized as follows.
Section~\ref{sec2} introduces the modeling framework.
Section~\ref{sec3} presents fluctuating-hydrodynamic simulations and perturbative analysis for Couette flow near solid walls.
Section~\ref{sec4} describes the MD-based protocols for determining $\eta_0$ and tests the obtained value in independent settings.
Section~\ref{sec5} discusses the role of the UV cutoff length.
Section~\ref{sec6} concludes the paper.

\section{Model}
\label{sec2}

\subsection{Microscopic and mesoscopic models}
\label{sec2-1}

At the microscopic level, we model the 2D fluid as a classical $N$-particle Hamiltonian system.
We use a repulsive soft-core potential
\begin{align}
    V(r) =
    \begin{cases}
        k (\sigma - r)^{\alpha} & r < \sigma, \\
        0 & r\geq \sigma.
    \end{cases}
\end{align}
Here, $k$ is the interaction strength, $\sigma$ is the particle diameter, and $\alpha$ is the exponent characterizing the steepness of the repulsion.

At the mesoscopic level, we assume that the system is described by fluctuating hydrodynamics for an isothermal 2D fluid.
The density field $\rho(\bm{r},t)$ and velocity field $\bm{v}(\bm{r},t)$ evolve as
\begin{align}
    \frac{\partial \rho(\bm{r},t)}{\partial t}  = - \nabla \cdot (\rho \bm{v}), \label{eq: fluctuating equation of continuity}
\end{align}
\begin{align}
    \rho \left[ \frac{\partial \bm{v}}{\partial t} + (\bm{v} \cdot \nabla) \bm{v} \right]
    &= - \nabla p + \eta_0 \nabla^2 \bm{v} + \zeta_0 \nabla(\nabla \cdot \bm{v})
    - \nabla \cdot \bm{\Pi}^{\mathrm{ran}},
    \label{eq: fluctuating Navier-Stokes equation}
\end{align}
where $p(\bm{r},t)$ is the pressure field.
The random stress tensor $\bm{\Pi}^{\mathrm{ran}}(\bm{r},t)$ is Gaussian white noise satisfying the fluctuation-dissipation relation
\begin{align}
    \big\langle \bm{\Pi}^{\mathrm{ran}}_{ab}(\bm{r},t)
    \bm{\Pi}^{\mathrm{ran}}_{cd}(\bm{r}',t') \big\rangle
    &= 2 k_B T \delta(\bm{r}-\bm{r}')\delta(t-t') \biggl[\eta_0
    \left(\delta_{ac}\delta_{bd}+\delta_{ad}\delta_{bc}\right)
    + \left(\zeta_0 - \eta_0 \right) \delta_{ab}\delta_{cd} \biggr],
    \label{eq:random stress tensor}
\end{align}
where $T$ is the temperature and $k_{B}$ is the Boltzmann constant.

Equation~(\ref{eq: fluctuating Navier-Stokes equation}) corresponds to the Navier--Stokes equation with an additional stochastic stress $\bm{\Pi}^{\mathrm{ran}}(\bm{r},t)$.
We then define the noise-averaged stress tensor from the momentum flux as~\footnote{
In fluctuating hydrodynamics, the stress is defined as an extension of the deterministic stress~\cite{Landau1959-eu}.
}
\begin{align}
    \langle \bm{\sigma}_{ab} \rangle
    = -\langle \bm{\Pi}_{ab} \rangle
    + \langle \rho \rangle \langle v_a \rangle \langle v_b \rangle .
\end{align}
with
\begin{align}
    \bm{\Pi}_{ab}
    &= \rho v_a v_b+ p \delta_{ab}
    - \eta_0 \left(\partial_a v_b+\partial_b v_a\right)
    - (\zeta_0 - \eta_0) \delta_{ab} \nabla\cdot\bm{v}
    + \bm{\Pi}^{\mathrm{ran}}_{ab}.
    \label{eq: momentum flux}
\end{align}

In this paper, we consider a dense fluid where density variations are small and focus on the bare shear viscosity $\eta_0$.
To this end, in the theoretical analysis, we consider the incompressible limit by isolating the transverse momentum dynamics.
In simulations, we solve the compressible equations with $p = c_T^2 \rho$ and set $c_T^2=1000$ and $\zeta_0 = 1$, since solving the incompressible equations with boundaries is computationally demanding.
As shown later, the agreement with MD results suggests that this choice of $\zeta_0$ does not affect the extraction of $\eta_0$.

\subsection{Hydrodynamic behavior at the macroscopic scale}
\label{sec2-2}

While fluid motion inherently exhibits thermal fluctuations, macroscopic experimental observations probe noise-averaged quantities.
The noise-averaged shear stress $\langle \bm{\sigma}_{xy} \rangle$ is related to the noise-averaged velocity $\langle \bm{v}\rangle$ as:
\begin{align}
\langle \bm{\sigma}_{xy} \rangle = \etaR \biggl(\frac{\partial \langle v^x\rangle}{\partial y} + \frac{\partial \langle v^y\rangle}{\partial x} \biggr).
\label{eq: Newton's law}
\end{align}
This defines the observed viscosity $\etaR$.
Importantly, $\etaR$ differs from the bare viscosity $\eta_0$~\cite{Pomeau1975-ll, Forster1976-ea, Forster1977-lr}.
As shown in MCT and RG analyses, the observed viscosity $\etaR$ is decomposed as
\begin{align}
    \etaR = \eta_0 + \delta \eta,
    \label{eq:decomposition of renormalized viscosity}
\end{align}
and $\delta \eta$ arises from the nonlinear coupling of hydrodynamic fluctuations.
Historically, $\delta \eta$ is referred to as the mode-coupling contribution or the renormalization correction.

In particular, the correction $\delta \eta$ has been explicitly evaluated in bulk fluids~\cite{Pomeau1975-ll, Forster1976-ea, Forster1977-lr}.
In two dimensions, $\delta \eta$ diverges logarithmically with system size~\cite{Alder1960-bh, Pomeau1975-ll, Forster1976-ea, Forster1977-lr, Isobe2008-xd}.
\begin{align}
    \delta \eta \sim \log L \quad \text{as} \quad L \to \infty.
\end{align}
Thus, $\etaR$ depends on system size, whereas $\eta_0$ is a system-size-independent quantity determined by microscopic interactions.
A brief review of this anomalous behavior is given in Appendix~\ref{app1}.

In the following, all quantities are expressed in reduced units based on the atomic system, where the atomic mass $m$, the atomic diameter $\sigma$, and the temperature $T$ are set to $1$.
We fix the mean density $\rho_0 = 0.765$ to focus on a dense fluid.
We also introduce a UV cutoff length $a_{\mathrm{uv}}$ as the unit of spatial discretization in fluctuating hydrodynamics.
The decomposition in Eq.~(\ref{eq:decomposition of renormalized viscosity}) depends on the UV cutoff length $a_{\mathrm{uv}}$ in $d \geq 2$~\cite{delamotte2012introduction}.
Except in the final section (Sec.~\ref{sec5}), we set $a_{\mathrm{uv}} = \sigma$, corresponding to the particle diameter, and refer to $\eta_0$ defined with this choice as the bare viscosity; see Sec.~\ref{sec5} for details.

\section{Key idea from fluctuating hydrodynamics}
\label{sec3}

In this section, using fluctuating hydrodynamics, we show that the bare viscosity $\eta_0$ manifests in observable quantities near solid walls.
This observation forms the basis of an operational method to determine $\eta_0$ from molecular dynamics.

\begin{figure}[t]
\begin{center}
\includegraphics[width=\textwidth]{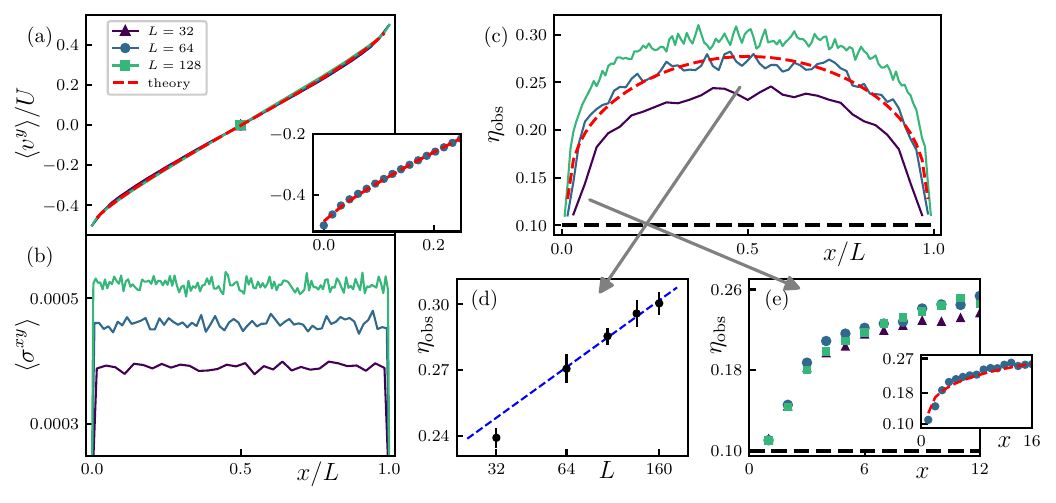}
\end{center}
\vspace{-0.5cm}
\caption{
Simulation and theoretical results for system-size-dependent observables in Couette flow within fluctuating hydrodynamics.
Colored lines: simulation results for $L=32$ (purple), $64$ (blue), and $128$ (green).
Red lines: theoretical results [Eqs.~(\ref{eq:velocity profile in pt}) and (\ref{eq:etaR in pt})].
The black dashed line in (c) and (e) represents the input value $\eta_0=0.10$.
(a) Scaled velocity profiles $\langle v^y(x)\rangle/U$ vs.\ $x/L$.
Inset: close-up near the wall.
(b) Shear stress profiles $\langle \sigma^{xy}(x)\rangle$ vs.\ $x/L$.
(c) Observed viscosity profiles $\eta_R(x)$ vs.\ $x/L$.
(d) System-size dependence of bulk $\eta_R$.
(e) Close-up of $\eta_R(x)$ near the wall.
Inset: comparison between simulation and theory.
Parameters: $\rho_0=0.765$, $T=1.0$, $\eta_0=0.10$, $U/L=0.002$, $\auv=1.0$, $c_T^2=1000$ (approximate incompressibility), and $\zeta_0=1.0$.
}
\label{fig1}
\end{figure}
\subsection{Numerical simulation of fluctuating hydrodynamics}
\label{sec3-1}

We consider Couette flow between two parallel walls at $x=0$ and $x=L$, moving with velocities $\mp U/2$ along the $y$ axis.
We perform direct numerical simulations of the fluctuating hydrodynamic equations [Eqs.~(\ref{eq: fluctuating equation of continuity}) and (\ref{eq: fluctuating Navier-Stokes equation})] with no-slip boundary conditions:
\begin{align}
v^x = 0 \quad &{\rm at} \ x=0,L, \label{eq:BC1}\\
v^y = \mp U/2 \quad &{\rm at} \ x=0,L,  \label{eq:BC2}\\
\partial_x \rho = 0 \quad &{\rm at} \ x=0,L.  \label{eq:BC3}
\end{align}
Periodic boundary conditions are imposed along $y$.
The numerical method is described in Appendix~\ref{app2}.

To disentangle the contributions of the bare viscosity and fluctuations, we use their distinct system-size dependence (see Sec.~\ref{sec2-2}).
Specifically, we perform simulations for varying system sizes and analyze the resulting scaling behavior.
The results are shown in Fig.~\ref{fig1}.
The simulations are conducted at a constant shear rate $U/L$ in the linear response regime (see Appendix~\ref{app3}).

Figure~\ref{fig1}(a) shows the scaled velocity profiles, $\langle v^y\rangle/U$, as a function of $x/L$.
As seen in the figure, the observed profile exhibits clear non-linear behavior, particularly near the walls.
The shear stress $\langle \sigma^{xy}(x)\rangle$ in Fig.~\ref{fig1}(b) is spatially uniform but exhibits a system-size dependence.
These behaviors significantly deviate from deterministic hydrodynamics, which predicts a purely linear velocity profile and a size-independent shear stress, suggesting the strong influence of thermal fluctuations.

The spatially uniform shear stress combined with the non-uniform velocity gradient implies that the observed viscosity becomes position dependent.
Figure~\ref{fig1}(c) shows the profiles of $\etaR(x)$ for different system sizes.
To quantify the bulk behavior, we extract the value of $\etaR$ in the bulk region and plot it as a function of $L$ in Fig.~\ref{fig1}(d).
The bulk value increases logarithmically with system size, consistent with previous predictions from MCT and RG analyses~\cite{Alder1960-bh, Pomeau1975-ll, Forster1976-ea, Forster1977-lr, Isobe2008-xd}.

Notably, near the walls, $\etaR(x)$ becomes independent of system size, exhibiting a scaling behavior distinct from that in the bulk.
This is shown in Fig.~\ref{fig1}(e), which presents a close-up near the boundaries.
The value of $\etaR(x)$ at the walls is nearly identical to $\eta_0 = 0.10$, the bare viscosity used in the simulations.
This indicates that the wall region isolates the bare contribution from fluctuation-induced effects.

We now discuss the physical origin of this behavior.
From a technical standpoint in our simulations, the no-slip boundary condition [Eqs.~(\ref{eq:BC1})-(\ref{eq:BC3})] effectively eliminates velocity fluctuations at the wall, leading to a strong reduction of the fluctuation-induced correction $\delta \eta$.
However, we emphasize that this suppression is not merely an artifact of the chosen boundary condition, but originates fundamentally from momentum scattering at the solid boundary.
In the bulk, fluctuation-induced corrections are driven by long-wavelength velocity fluctuations, which rely on local momentum conservation.
At the boundary, in contrast, collisions with the wall allow fluid momentum exchange with the solid boundary and suppress the long-wavelength fluctuations responsible for the mode-coupling effect.
This suppression of macroscopic fluctuations is the fundamental cause of the observed behavior.
Consistent with this picture, a similar suppression of fluctuations is observed in the MD simulations presented in Sec.~\ref{sec4}.

\subsection{Analytical expressions for the velocity profile and local viscosity}
\label{sec3-2}

We here develop a perturbative theory to derive analytical expressions for the quantities observed in the simulations.
We consider an incompressible fluid with boundary conditions corresponding to Eqs.~(\ref{eq:BC1})–(\ref{eq:BC3}); details are given in Appendix~\ref{app4}.
Below we present the results.

First, the mean velocity is along the $y$ axis and depends only on $x$.
The perturbative calculation gives
\begin{align}
   \langle v^y(\bm{r}) \rangle = \dot{\gamma} \left[ x + \frac{A}{L} \sum_{k_x}\frac{1}{k_x}\frac{\sin(2k_x x)}{2k_x} \right] \label{eq:velocity profile in pt}
\end{align}
with
\begin{align}
    A = \frac{\rho_0 k_BT}{4\eta_0^2}, \label{eq:numerical factor in pt}
\end{align}
where $\dg:=U/L$, $k_x := (\pi/L)n_x$ and the summation is carried out from $n_x=1$ to $n_x=n_x^{\rm max} := L/\auv$.
Force balance makes the shear stress $\langle \sigma^{xy}(\bm{r}) \rangle$ independent of $\bm{r}$:
\begin{align}
    \langle \sigma^{xy} \rangle &= \eta_0 \dot{\gamma} \biggl[1 + \frac{A}{L} \sum_{k_x}\frac{1}{k_x}\biggr].
    \label{eq:shear stress profile in pt}
\end{align}
Combining the shear stress Eq.~(\ref{eq:shear stress profile in pt}) and the velocity profile Eq.~(\ref{eq:velocity profile in pt}), the observed viscosity is analytically given as
\begin{align}
     \etaR(x) &= \eta_0 \biggl[ 1 + 2\frac{A}{L} \sum_{k_x}\frac{1}{k_x}\sin^2(k_xx)\biggr].
    \label{eq:etaR in pt}
\end{align}
This equation provides the explicit expression of the decomposition in Eq.~(\ref{eq:decomposition of renormalized viscosity}).
The second terms in Eqs.~(\ref{eq:velocity profile in pt}), (\ref{eq:shear stress profile in pt}), and (\ref{eq:etaR in pt}) are fluctuation-induced contributions, which lead to spatially non-uniform velocity gradients and viscosity.
In particular, at the midpoint $x=L/2$, the observed viscosity is calculated as:
\begin{align}
    \etaR(x=L/2)&\sim \eta_0 \biggl(1+ \frac{A}{\pi}\int^{2\pi/\auv}_{2\pi/L} dk_x \frac{1}{k_x} \biggr)\\
    &=\eta_0\biggl[1+ \frac{A}{\pi} \log\biggl(\frac{L}{\auv}\biggr) \biggr].
    \label{eq: renor_visc at bulk}
\end{align}
Thus, $\delta \eta(x = L/2) := \etaR(x = L/2) - \eta_0$ diverges logarithmically as $L \to \infty$.
In addition, the bare viscosity is directly observed at the walls:
\begin{align}
\etaR(x = 0) = \etaR(x = L) = \eta_0.
\label{eq: renor_visc at wall}
\end{align}
Equations~(\ref{eq:etaR in pt}), (\ref{eq: renor_visc at bulk}), and (\ref{eq: renor_visc at wall}) provide mathematical confirmation of our simulation results and support the central observation that the near-wall noise-averaged behavior is governed by the bare viscosity.

We check the validity of Eqs.~(\ref{eq:velocity profile in pt})–(\ref{eq:etaR in pt}).
Revisiting Fig.~\ref{fig1}(a), we compare the velocity field with the theoretical prediction.
To this end, we fit Eq.~(\ref{eq:velocity profile in pt}) to the simulation data for $L=64$, using $\dot{\gamma}$ and $A$ as fitting parameters.
The theoretical expression is in good agreement with the numerical result.
The fitted value of $A$ does not exactly match the theoretical prediction [Eq.~(\ref{eq:numerical factor in pt})], reflecting the approximations made in the perturbative analysis.

A similar comparison for the local viscosity is shown in Fig.~\ref{fig1}(c) and the inset of (e).
Fitting Eq.~(\ref{eq:etaR in pt}) with $\eta_0$ and $A$ again yields excellent agreement with the simulation results, even near the walls.
Despite the discrepancy in $A$, the fitted value of $\eta_0 = 0.128$ remains close to the simulation value $\eta_0 = 0.100$, indicating that the near-wall structure of $\eta_R(x)$ enables a robust extraction of $\eta_0$.

\begin{figure}[t]
\begin{center}
\includegraphics[width=\textwidth]{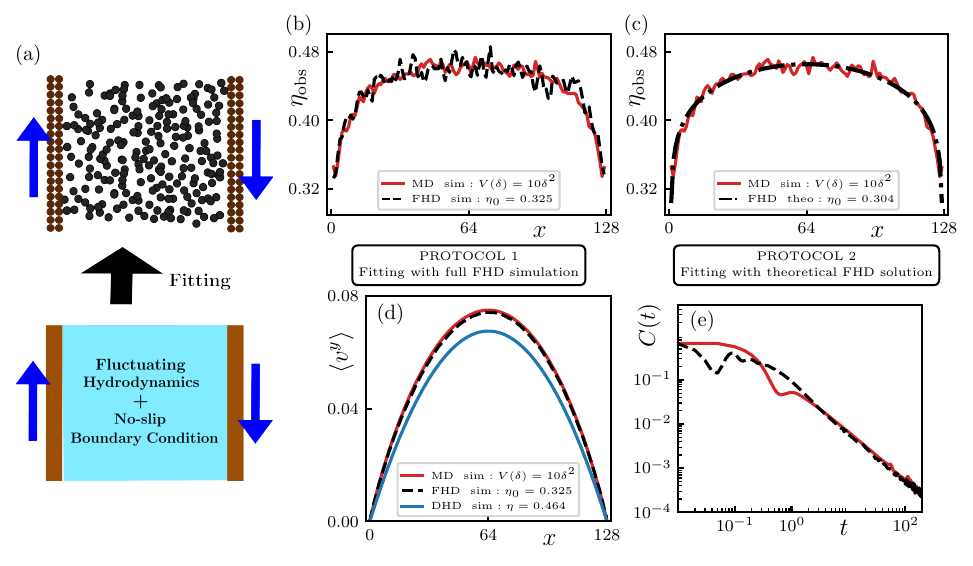}
\end{center}
\vspace{-0.5cm}
\caption{
Measurement protocols for determining the bare viscosity $\eta_0$ and their validation.
(a) Schematic of the method focusing on near-wall behavior in Couette flow.
(b) Protocol 1: $\etaR(x)$ from MD (red) fitted by fluctuating hydrodynamics (black), yielding $\eta_0=0.325$.
(c) Protocol 2: $\etaR(x)$ (red) fitted by the analytical expression [Eq.~(\ref{eq:etaR in pt})] (black), yielding $\eta_0=0.304$.
(d) Validation in Poiseuille flow: velocity profile from MD (red) compared with fluctuating hydrodynamics (black, using $\eta_0=0.325$) and deterministic hydrodynamics (blue, using bulk $\etaR=0.464$).
(e) Validation via equilibrium momentum density correlation: MD (red) vs fluctuating hydrodynamics (black, using $\eta_0=0.325$).
Parameters: $V(\delta)=10\delta^2$, $\rho_0=0.765$, $T=1.0$, $L=128.0$.
Flow conditions: (b,c) Couette flow with $U/L=0.0014$; (d) Poiseuille flow with an external force $g=0.00002$.
}
\label{fig2}
\end{figure}
\section{Measurement protocols of the bare viscosity in molecular dynamics}
\label{sec4}

Building on the results of Sec.~\ref{sec3}, we develop operational protocols to measure the bare viscosity in molecular dynamics.

\subsection{Setup of atomic systems}
\label{sec4-1}

To determine the bare viscosity from the atomic description, we perform MD simulations of the atomic system introduced in Sec.~\ref{sec2-1}.
Our approach involves fitting the near-wall behavior of the fluid in the Couette geometry using fluctuating hydrodynamics.
This method is illustrated in Fig.~\ref{fig2}(a).

In our MD simulations, the walls are modeled as particles arranged on a square lattice.
They are confined by an onsite potential and thermostatted via a Langevin thermostat to maintain the temperature.
Wall–fluid interactions are taken to be purely repulsive in the main text, mimicking a hydrophobic surface.
Details of the interaction potentials and simulation parameters are provided in Appendix~\ref{app5}, and we confirm in Appendix~\ref{app6} that the results are robust for other types of microscopic walls.

\subsection{Measurement protocol of bare viscosity}
\label{sec4-2}

The protocol is based on the theoretical result that the fluctuation-induced correction vanishes at the wall and remains strongly suppressed in the near-wall region (Sec.~\ref{sec3}).
The concrete procedure consists of the following steps:
\newline
\newline
\textbf{Protocol 1}
\begin{enumerate}
    \item Measure $\etaR(x)$ in atomic systems.
    \item Fit obtained $\etaR(x)$ with the results of fluctuating hydrodynamics simulations.
\end{enumerate}

Figure~\ref{fig2}(b) shows a fitting result for the interparticle potential $V(\delta)=10\delta^2$.
To ensure dimensional consistency, we match $\rho_0$, $k_B T$, and $L$ between the two descriptions and treat $\eta_0$ as the only fitting parameter.
In practice, $\eta_0$ is varied in increments of $0.005$ to identify the value that best reproduces the data.
As shown in this figure, the best-fitted curve accurately reproduces the MD data near the wall, yielding a bare viscosity of $\eta_0 = 0.325$ for this fluid.
This agreement supports the interpretation that the fitted parameter corresponds to the bare viscosity.

While this protocol provides an accurate estimate of $\eta_0$, numerically solving the fluctuating hydrodynamic equations can be computationally demanding.
To address this limitation, we introduce a simplified protocol based on the analytical expression derived in Sec.~\ref{sec3-2}.
\newline
\newline
\textbf{Protocol 2}
\begin{enumerate}
    \item Measure $\etaR(x)$ in atomic systems.
    \item Fit the obtained $\etaR(x)$ to the analytical expression Eq.~(\ref{eq:etaR in pt}), using $\eta_0$ and $A$ as fitting parameters.
\end{enumerate}

Figure~\ref{fig2}(c) shows the application of Protocol 2 to the same atomic system as Protocol 1.
The analytical expression Eq.~(\ref{eq:etaR in pt}) with the fitted $\eta_0$ (black line) closely follows the MD data (red circles).
Protocol 2 yields $\eta_0 = 0.304$, consistent with the value obtained from Protocol 1.

We further apply both protocols to other atomic systems.
As shown in Appendix~\ref{app7}, we confirm that this approach is successful for all the systems investigated.
As summarized in Table~\ref{tab1}, the value of $\eta_0$ obtained from Protocol 2 is quantitatively consistent with that from Protocol 1 across different systems, validating this computationally efficient approach.
However, since Eq.~(\ref{eq:etaR in pt}) is derived from perturbation theory, small deviations from the more accurate Protocol 1 values can arise.
The accuracy of Protocol 2 improves as $\delta\eta/\eta_0$ decreases, consistent with the perturbative nature of Eq.~(\ref{eq:etaR in pt}).

\subsection{Validity of our protocols}
To verify that the extracted $\eta_0$ is an intrinsic material constant rather than an artifact of the Couette geometry, we test its predictive capability in different setups.
We first consider Poiseuille flow, where the fluid is confined between two fixed parallel walls and driven by a constant body force.
Figure~\ref{fig2}(d) shows the velocity profile obtained from MD simulations, together with the prediction of fluctuating hydrodynamics using the previously extracted $\eta_0$.
We observe that the fluctuating hydrodynamics prediction accurately reproduces the MD results, which demonstrates that the extracted $\eta_0$ consistently describes different flow geometries.
This successful agreement also implies that one can alternatively extract $\eta_0$ directly from Poiseuille flow data using fluctuating hydrodynamics; such an analysis is presented in Appendix~\ref{app8}.

For comparison, we also show the prediction of deterministic hydrodynamics, using the viscosity measured in the bulk region of the Couette flow.
This fails to reproduce the MD profile, highlighting the necessity of fluctuating hydrodynamics near the wall.

\begin{table}[tb]
\centering
\begin{tabularx}{0.8\columnwidth}{@{\extracolsep{\fill}}c|c|c}
\hline
Atomic System & Protocol 1 & Protocol 2 \\
\hline
$\quad V(\delta) = 10 \delta^2 \quad$ & 0.325 & 0.304\\
$\quad V(\delta) = 10 \delta^4 \quad$ & 0.470 & 0.450 \\
$\quad V(\delta) = 10 \delta^6 \quad$ & 0.660 & 0.642 \\
\hline
\end{tabularx}
\caption{Estimation of bare viscosity for the three atomic systems.
The results for the atomic system with $V(\delta) = 10 \delta^2$ are shown in Fig.~\ref{fig2}.
}
\label{tab1}
\end{table}

As a further consistency check, we examine equilibrium fluctuations of the fluid~\cite{Alder1960-bh, Dorfman1975-pr, Pomeau1975-ll, Lowe1995-am, Isobe2008-xd}.
We consider the time correlation function of the momentum density field $\bm{j} := \rho \bm{v}$, defined as
\begin{align}
C_{\rm eq}(t) := \frac{1}{2S_{\mathcal{B}}} \int_{\mathcal{B}} d^2\bm{r} \langle \bm{j}(\bm{r},t) \cdot \bm{j}(\bm{r},0)\rangle_{\rm eq},
\label{eq: definition of time correlation function of the momentum field}
\end{align}
where $\mathcal{B} = [L/4, 3L/4] \times [L/4, 3L/4]$ is chosen to minimize boundary effects and $S_{\mathcal{B}} = L^2/4$.
Figure~\ref{fig2}(e) shows that $C_{\rm eq}(t)$ obtained from MD simulations is accurately reproduced by fluctuating hydrodynamics with the $\eta_0$ extracted from the Couette flow.
This provides independent validation of our approach.

We emphasize that the agreement between fluctuating hydrodynamics and MD simulations persists down to microscopic scales.
As shown in Fig.~\ref{fig2}(e), the time correlation function $C_{\rm eq}(t)$ is accurately reproduced even at short times ($t \geq 10$).
Similarly, for steady Couette and Poiseuille flows [Figs.~\ref{fig2}(b) and (d)], fluctuating hydrodynamics reproduces the MD profiles with spatial resolution on the order of the atomic diameter, including the near-wall region.
These results demonstrate the applicability of fluctuating hydrodynamics at microscopic scales, despite being a continuum theory.

\begin{figure}[tb]
\begin{center}
\includegraphics[scale=1]{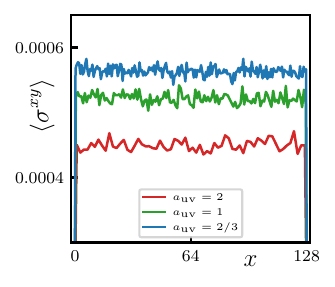}
\end{center}
\vspace{-0.5cm}
\caption{
Influence of the UV cutoff $\auv$ on the shear stress profile in fluctuating hydrodynamics simulations.
Shear stress profiles $\langle \sigma^{xy} \rangle$ are shown as a function of position $x$ for different values of $\auv$ with a fixed input viscosity $\eta_0$.
The different colors represent different $\auv$ values: $\auv=2$ (red), $\auv=1$ (green), and $\auv=2/3$ (blue), all with the same input viscosity $\eta_0=0.1$.
Parameters are the same as in Fig.~\ref{fig1}.
}
\label{fig3}
\end{figure}
\begin{figure}[tb]
\begin{center}
\includegraphics[width=\textwidth]{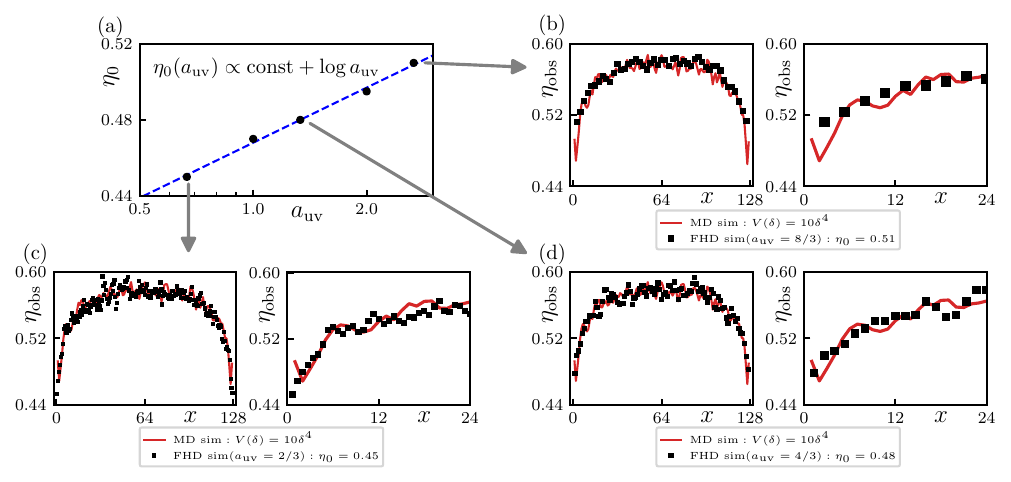}
\end{center}
\vspace{-0.5cm}
\caption{
Dependence of the input viscosity $\eta_0$ on the UV cutoff $\auv$ and its effect on the accuracy of fluctuating hydrodynamics.
(a) Relationship between $\auv$ and the corresponding optimal input viscosity $\eta_0$ for the atomic system, determined using Protocol 1.
The parameters for the atomic system are $V(\delta) = 10 \delta^4$, $\rho_0=0.765$, $T=1.0$, and $L=128.0$.
The range of $\auv$ explored is $2/3 \leq \auv \leq 8/3$.
(b)-(d) Comparison of MD simulation results (red lines) and fluctuating hydrodynamics predictions (black dots) for three representative values of $2/3 \leq \auv \leq 8/3$.
Each figure consists of two panels: the left-hand side panel showing the overall view, and the right-hand side panel presenting a zoomed-in view near the wall.
}
\label{fig4}
\end{figure}
\section{Role of the UV cutoff length on fluctuating hydrodynamics}
\label{sec5}

Fluctuating hydrodynamics describes fluid motion coarse-grained over a length scale larger than the UV cutoff $\auv$.
While $\auv$ was fixed to the atomic diameter $\sigma$ in the previous sections, it should be regarded as a hydrodynamic parameter absent in the microscopic MD description.
We therefore investigate the impact of varying $\auv$ in this section.
In the following, $\eta_0$ denotes the input viscosity in the fluctuating hydrodynamic framework.

We first examine the effect of $\auv$ at fixed input viscosity $\eta_0$.
Figure~\ref{fig3} shows that the shear stress $\langle \sigma^{xy} \rangle$ increases as $\auv$ decreases.
This demonstrates that $\auv$ directly affects macroscopic observables by controlling the range of fluctuations included in the description.

The framework of fluctuating hydrodynamics resembles that of quantum field theory rather than deterministic hydrodynamics.
Deterministic hydrodynamics lacks fluctuations and renormalization effects, and is therefore independent of $\auv$.
In contrast, fluctuating hydrodynamics and quantum field theory include thermal and quantum fluctuations, respectively, leading to renormalization effects.
In quantum field theory, Wilson's RG theory provides a systematic way to change the scale $\Lambda := 2\pi / \auv$~\cite{polchinski1992effective, kaplan5lectures, 10.1143/PTPS.131.395, ROSTEN2012177}.
When $\Lambda$ is changed and the parameters are tuned along the RG flow, the model describes the same physical system at different resolutions.

Similarly, in fluctuating hydrodynamics, the input viscosity becomes a scale-dependent quantity $\eta_0(\auv)$, which provides the appropriate hydrodynamic description for a given cutoff $\auv$.
To demonstrate this, we consider the atomic system with the interparticle potential $V(\delta)=10\delta^4$.
We vary $\auv$ within $2/3 \leq \auv \leq 8/3$ and determine the optimal $\eta_0$ for each value using Protocol 1.
The resulting relation $\eta_0(\auv)$ is shown in Fig.~\ref{fig4}(a), and Figs.~\ref{fig4}(b)–(d) confirm that fluctuating hydrodynamics with these parameter pairs accurately reproduces the MD results.

As $\auv$ increases, fine-scale structures near the wall are progressively smoothed out.
Nevertheless, the agreement with MD simulations remains robust.
These results indicate that $\eta_0(\auv)$ plays a role analogous to an RG flow.
Importantly, this flow is obtained here through direct numerical simulations, rather than through an explicit RG calculation.

Finally, we examine the behavior of $\eta_0(\auv)$ at small $\auv$.
The numerical results are well fitted by a logarithmic form, $\eta_0(\auv)=C_0+C_1\log\auv$, indicating a logarithmic divergence as $\auv \to 0$ (see Appendix~\ref{app1}).
Such ultraviolet (UV) divergences are also common in quantum field theory.
In fluctuating hydrodynamics, however, this divergence is not problematic because the description is valid only above a minimum length scale $\abare$.
This lower bound reflects the breakdown of the hydrodynamic description below microscopic scales.
We therefore define the bare viscosity as $\eta_{\rm bare} := \eta_0(\abare)$~\footnote{In quantum field theory, the term {\it bare parameter} typically refers to the parameter in the continuum limit, such as $\lim_{\auv \to 0} \eta_0(\auv)$, which diverges.}.

Conventional arguments often associate $\abare$ with the mean free path.
In the dense fluids considered here, the mean free path is comparable to the atomic diameter $\sigma$.
Our numerical results (Figs.~\ref{fig2} and \ref{fig4}) demonstrate that fluctuating hydrodynamics remains valid down to this scale.
Moreover, the logarithmic dependence of $\eta_0(\auv)$ on $\auv$ implies weak sensitivity to the precise choice of the cutoff.
Taken together, we suggest that the bare scale can be identified with the atomic diameter, $\abare \simeq \sigma$, and thus $\eta_{\rm bare} = \eta_0(\sigma)$.
Importantly, we emphasize that this identification is not an assumption but a direct numerical finding of the present study.
Thus, throughout this paper, we understood $\eta_0=\eta_0(\sigma)$ as the bare viscosity.

This identification should be understood within the steady-state settings studied here.
In general, the bare scale $\abare$ may depend on the physical process and could differ, for example, in time-dependent flows.
The identification $\abare \simeq \sigma$ in steady states may be related to the fact that time averaging effectively coarse-grains stochastic atomic motion.
Establishing a general criterion for determining $\abare$ remains an important open problem for future research.

\section{Discussions and concluding remarks}
\label{sec6}

In this study, we developed operational protocols for determining the bare viscosity $\eta_0$ of fluctuating hydrodynamics from microscopic dynamics.
The central idea is that solid boundaries suppress the hydrodynamic fluctuations responsible for renormalizing the viscosity in the bulk.
As a result, the near-wall response retains direct information on $\eta_0$.
We demonstrated this mechanism using direct numerical simulations of fluctuating hydrodynamics.
We then used this near-wall behavior as the basis for determining $\eta_0$ in MD simulations.
From the noise-averaged one-body observables in Couette flow, we extracted $\eta_0$ and confirmed that the same value describes Poiseuille flow and equilibrium momentum-density correlations.

We also clarified the role of the UV cutoff length $\auv$ in fluctuating hydrodynamics.
Changing $\auv$ changes the input viscosity required to reproduce the same atomic system, so that $\eta_0(\auv)$ plays the role of an RG flow.
For the steady states studied here, fluctuating hydrodynamics remains quantitatively accurate down to $\auv$ of the order of the atomic diameter.
This supports the identification of the bare scale as $\abare \simeq \sigma$ and the bare viscosity as $\eta_{\rm bare}=\eta_0(\sigma)$.
Whether the same identification holds for time-dependent phenomena remains an open question.

\subsection{Three-dimensional systems}
Although fluctuation-induced corrections are generally smaller in three dimensions, the distinction between the bare viscosity in fluctuating hydrodynamics and the macroscopic viscosity in deterministic hydrodynamics remains conceptually important.
The boundary-based protocol developed here can also be applied to three-dimensional fluids.
For example, using the perturbation theory developed in Sec.~\ref{sec3-2} and Appendix~\ref{app4}, the observed viscosity in three dimensions is obtained as
\begin{align}
     \etaR(x) &= \eta_0 \biggl[ 1 + 2\frac{A}{L} \sum_{k_x}\mathrm{arcsinh}\biggl(\frac{\Lambda}{k_x}\biggr) \sin^2\biggl(k_xx\biggr)\biggr]
    \label{eq:etaR in pt: 3d}
\end{align}
with $\Lambda = 2 \pi / \auv$.
The fluctuation-induced term again vanishes at the wall because of the factor $\sin^2(k_xx)$.
Thus, high-resolution measurements near solid boundaries should provide access to $\eta_0$ in three-dimensional fluids as well.

\subsection{Experimental realization}
We finally discuss possible experimental realizations of the present protocol.
Classical examples are soap films and free-standing smectic liquid-crystal films, which have been used to study two-dimensional turbulence~\cite{Rivera2000-xi, Kellay2002-ig} and the Stokes paradox~\cite{Eremin2011-ft}.
Vacuum-suspended liquid-crystal mono- or bilayers are particularly relevant because friction with the surrounding medium is strongly reduced, making them closer to an isolated two-dimensional fluid~\cite{Brogioli2017-hf}.

Electron fluids provide another class of two-dimensional hydrodynamic systems.
Hydrodynamic transport has been reported in graphene and related clean materials~\cite{Gooth2018-bs, Sulpizio2019-bq, Polini2020-gr, Fritz2024-yb}, and boundary scattering can be controlled in mesoscopic devices~\cite{Keser2021-rn}.
For both classical films and electron fluids, improved control of boundary conditions may make it possible to test the fluctuation-suppression mechanism discussed in this paper.
Although such experimental applications remain a future challenge, the present protocol already provides a direct way to determine bare transport coefficients in MD simulations.

Finally, the boundary-based idea is not limited to viscosity.
In one- and two-dimensional heat-conduction systems, divergent thermal conductivities have been observed in carbon nanotubes and graphene~\cite{Chang2008-zw, Xu2014-oo, Lee2017-gq, yang2021observation}.
Extending the present approach to bare thermal conductivity is a possible direction for future work.

The code used to obtain the findings in this paper and the data obtained from our simulation can be found in~\footnote{\url{https://isspns-gitlab.issp.u-tokyo.ac.jp/nakano35255/comp_fluct_hydro} (in preparation)}.

\begin{acknowledgments}
We thank K. Yokota for their helpful comments.
The computation in this study has been done using the facilities of the Supercomputer Center, the Institute for Solid State Physics, the University of Tokyo.
The authors are grateful for the fruitful discussions at the workshop “Advances in Fluctuating Hydrodynamics: Bridging the Micro and Macro Scales” hosted by YITP at Kyoto University and RIKEN iTHEMS.
HN is supported by JSPS KAKENHI Grants No.JP22K13978.
YM is supported by the Ogawa science and technology foundation.
KS is supported by JSPS KAKENHI Grant No.JP23K25796.
\end{acknowledgments}

\clearpage
\appendix

\section{Properties of observed viscosity in bulk region}
\label{app1}

This appendix summarizes the bulk behavior of the observed viscosity $\etaR$ from the standard long-time-tail perspective.
In contrast to the boundary-driven shear flows considered in the main text, we here consider an equilibrium fluid in a periodic domain and recall how hydrodynamic fluctuations renormalize the viscosity measured by the Green--Kubo formula.

We consider a $d$-dimensional periodic box of linear size $L$.
The observed shear viscosity is given by the Green--Kubo relation
\begin{align}
    \etaR = \frac{L^d}{k_B T} \int_{0}^{\infty} ds \left\langle \bm{\pi}_{xy}(s) \bm{\pi}_{xy}(0) \right\rangle_{\rm eq}.
    \label{eq:green-kubo formula}
\end{align}
Here, $\bm{\pi}_{xy}(s)$ denotes the spatial average of the $xy$ component of the momentum flux,
\begin{align}
    \bm{\pi}_{xy}(s) := \frac{1}{L^d} \int d^d\bm{r} \bm{\Pi}_{xy}(\bm{r}, s),
\end{align}
where $\bm{\Pi}_{xy}(\bm{r}, s)$ is given by Eq.~(\ref{eq: momentum flux}).

Linearized fluctuating hydrodynamics predicts three regimes for the equilibrium stress correlation in Eq.~(\ref{eq:green-kubo formula}) \cite{Pomeau1975-ll, Atzberger2006-zo}:
\begin{equation}
    \left\langle \bm{\pi}_{xy}(t) \bm{\pi}_{xy}(0) \right\rangle_{\rm eq}
    \sim
    \begin{cases}
        2\eta_0 L^{-d} k_B T \delta(t),
        & 0 < t \ll \auv^2 / \eta_0,  \\[4pt]
        C_{\rm LT} L^{-d} t^{-d/2},
        & \auv^2 / \eta_0 \ll t \ll L^2 / \eta_0, \\[4pt]
        0,
        & L^2 / \eta_0 \ll t,
    \end{cases}
    \label{eq:stress_auto_correlation}
\end{equation}
where $C_{\rm LT}$ is the amplitude of the long-time tail.
The short-time delta function follows from the fluctuation-dissipation relation for the random stress [Eq.~(\ref{eq:random stress tensor})].
The intermediate regime is the long-time tail, $t^{-d/2}$, which originates from the slow relaxation of hydrodynamic modes.
This tail is also observed in our equilibrium simulations [Fig.~\ref{fig2}(e)].
Substituting Eq.~(\ref{eq:stress_auto_correlation}) into the Green--Kubo relation gives the leading correction to $\etaR$.
For $d=2$, we obtain
\begin{align}
    \etaR
    &\simeq \eta_0
    + \frac{2C_{\rm LT}}{k_B T}\log (L/\auv).
    \label{eq:etaMCT2d}
\end{align}
For $d>2$, the corresponding result is
\begin{align}
    \etaR
    &\simeq \eta_0
    + \frac{2C_{\rm LT}}{(d-2)k_B T}
    \biggl[
    \left(\frac{\eta_0}{\auv^2}\right)^{d/2-1}
    -
    \left(\frac{\eta_0}{L^2}\right)^{d/2-1}
    \biggr].
    \label{eq:etaMCTdgt2}
\end{align}

These estimates capture the leading dependence of $\etaR$ on the system size $L$ and the UV cutoff $\auv$.
In two dimensions, Eq.~(\ref{eq:etaMCT2d}) shows the logarithmic growth of $\etaR$ with $L$, consistent with our fluctuating hydrodynamic simulations [Figs.~\ref{fig1}(e) and \ref{fig3}].
For $d>2$, Eq.~(\ref{eq:etaMCTdgt2}) still contains a finite-size correction, but $\etaR$ converges to a finite value as $L \to \infty$.
This limiting value corresponds to the viscosity used in macroscopic hydrodynamics.

Equations~(\ref{eq:etaMCT2d}) and (\ref{eq:etaMCTdgt2}) also show that $\etaR$ depends on the UV cutoff $\auv$.
As $\auv \to 0^+$, the correction diverges logarithmically in two dimensions and algebraically for $d>2$.
This cutoff dependence corresponds to an ultraviolet (UV) divergence in the field-theoretic sense.
It is not a physical pathology, because fluctuating hydrodynamics is an effective coarse-grained theory defined only above a microscopic cutoff scale.
As discussed in Sec.~\ref{sec5}, the cutoff dependence should instead be interpreted as the scale dependence of the bare transport coefficient.
For a general discussion of UV divergences and dimensionality in field theory, see Ref.~\cite{polchinski1992effective, delamotte2012introduction}.

\begin{figure}[tb]
\begin{center}
\includegraphics[scale=0.5]{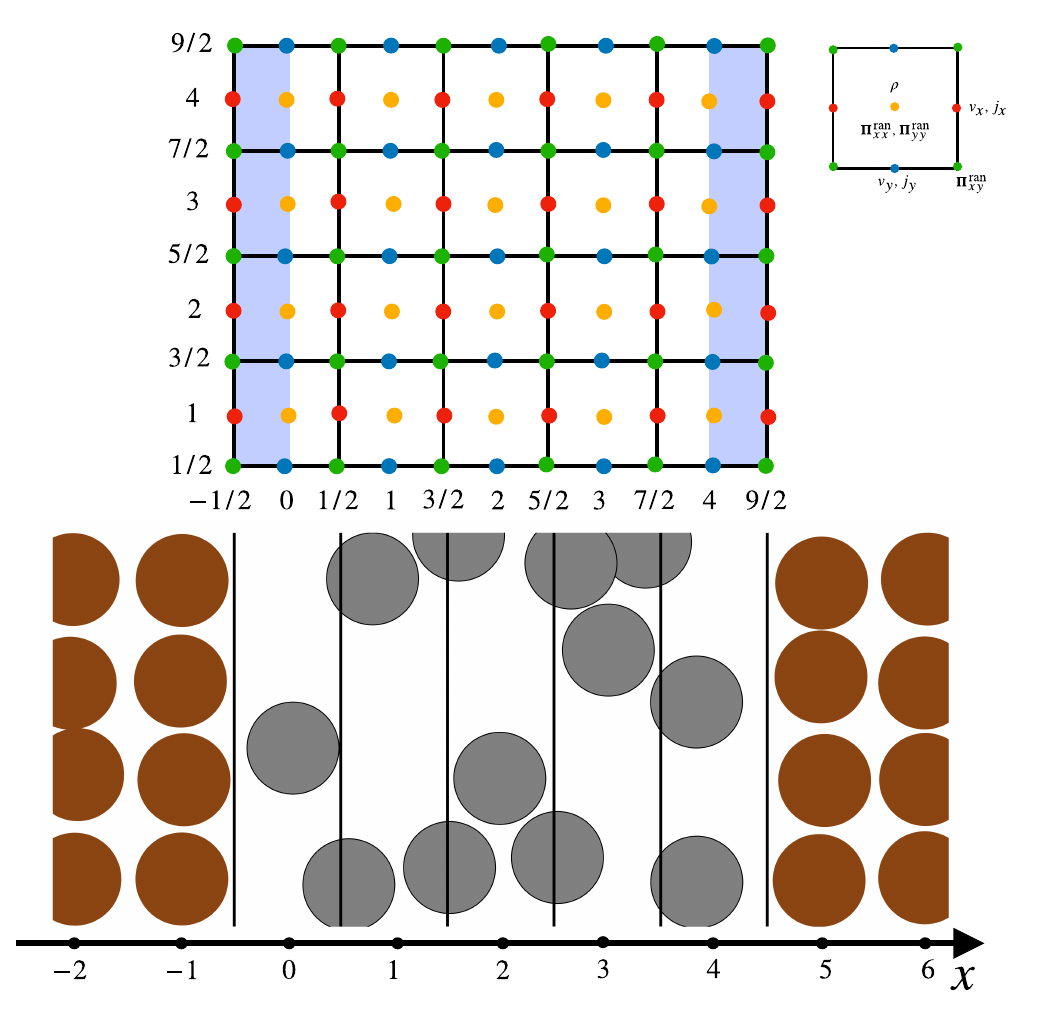}
\end{center}
\caption{
{Comparison of simulation boxes for fluctuating hydrodynamics and atomic system, using the case of $L = 4$ as an example.
(a) Fluctuating hydrodynamics simulation: The staggered lattice scheme is used, where different physical quantities are defined at distinct grid locations.
Boundary conditions are imposed at $x = 0$ and $x = 4$.
(b) MD simulation: Solid wall particles are placed at $x = -1$, $-2$, $-3$ and $x = 5$, $6$, $7$.
Physical quantities such as density and velocity fields are averaged over the regions $x \in [-0.5, 0.5]$, $[0.5, 1.5]$, $[1.5, 2.5]$, $[2.5, 3.5]$, and $[3.5, 4.5]$, and are considered to be located at $x = 0$, $1$, $2$, $3$, and $4$, respectively.
}
}
\label{supfig1}
\end{figure}

\section{Numerical solver of fluctuating hydrodynamics}
\label{app2}

o numerically solve the fluctuating hydrodynamic equations
[Eqs.~(\ref{eq: fluctuating equation of continuity}) and
(\ref{eq: fluctuating Navier-Stokes equation})], we use the staggered-grid
scheme developed in  Refs.~\cite{Garcia1987-fs, Mansour1987-im, Bell2007-yb, Voulgarakis2009-fl, Bell2010-ly, Donev2010-ew, BalboaUsabiaga2012-sh, Delong2013-fh, Balakrishnan2014-vg, Kim2017-cg, Narayanan2018-fr, Barker2023-ua, Srivastava2023-nx}.
This appendix describes the implementation used in our simulations.

\subsection{Equations used in the solver}
For the numerical implementation, we rewrite the equations in terms of the density $\rho$ and momentum density $\bm{j}:=\rho\bm{v}$:
\begin{align}
    \frac{\partial \rho}{\partial t} &= - \nabla \cdot \bm{j}, \\
    \frac{\partial j_a}{\partial t} &= - \bm{\nabla} \cdot (j_a \bm{v}) - \bm{\nabla}_a p(\rho)
    + \eta_0 \bm{\nabla}^2 v_a + \zeta_0 \bm{\nabla}_a (\bm{\nabla}\cdot \bm{v}) + \bm{\nabla}\cdot \bm{\Pi}^{\mathrm{ran}}_a,
\end{align}
with the equation of state
\begin{align}
p(\rho) &= c_T^2 \rho.
\end{align}
The random stress satisfying Eq.~(\ref{eq:random stress tensor}) is generated as
\begin{align}
    \bm{\Pi}^{\mathrm{ran}}_{xx} &= \sqrt{2k_B T \zeta_0} R_1 + \sqrt{2k_B T \eta_0} R_2, \\
    \bm{\Pi}^{\mathrm{ran}}_{yy} &= \sqrt{2k_B T \zeta_0} R_1 - \sqrt{2k_B T \eta_0} R_2, \\
    \bm{\Pi}^{\mathrm{ran}}_{xy} &= \bm{\Pi}^{\mathrm{ran}}_{yx} = \sqrt{2k_B T \eta_0} R_3,
\end{align}
where $R_a$ $(a=1,2,3)$ are independent Gaussian white noises with
\begin{align}
    \langle R_a(\bm{r}, t) R_b(\bm{r}', t')\rangle = \delta_{ab} \delta(\bm{r}-\bm{r}')\delta(t-t').
\end{align}

\subsection{Spatial discretization on a staggered grid}
The variables are placed on the staggered grid shown in Fig.~\ref{supfig1}:

\begin{description}
\item[Scalars] The density $\rho$ and pressure $p$ are defined at cell centers $(i,j)$.
\item[Vectors] The $x$ components $j_x$ and $v_x$ are defined at $x$ faces $(i+1/2,j)$, and the $y$ components $j_y$ and $v_y$ at $y$ faces $(i,j+1/2)$.
\item[Tensors] The diagonal random stresses $\bm{\Pi}^{\mathrm{ran}}_{xx}$ and $\bm{\Pi}^{\mathrm{ran}}_{yy}$ are defined at cell centers, while $\bm{\Pi}^{\mathrm{ran}}_{xy}$ is defined at cell edges $(i+1/2,j+1/2)$.
\end{description}

The momentum density at cell faces is evaluated by interpolation:
\begin{align}
    (j_x)_{i+1/2,j} &= \frac{\rho_{i,j}+\rho_{i+1,j}}{2}(v_x)_{i+1/2,j}, \\
    (j_y)_{i,j+1/2} &= \frac{\rho_{i,j}+\rho_{i,j+1}}{2}(v_y)_{i,j+1/2}.
\end{align}
The advective flux $j_a v_b$ is evaluated as
\begin{align}
    (j_x v_x)_{i,j}
    &= \frac{1}{4}
    \left[(j_x)_{i-1/2,j}+(j_x)_{i+1/2,j}\right]
    \times
    \left[(v_x)_{i-1/2,j}+(v_x)_{i+1/2,j}\right], \\
    (j_x v_y)_{i,j}
    &= \frac{1}{4}
    \left[(j_x)_{i-1/2,j}+(j_x)_{i+1/2,j}\right]
    \times
    \left[(v_y)_{i,j-1/2}+(v_y)_{i,j+1/2}\right], \\
    (j_y v_x)_{i,j}
    &= \frac{1}{4}
    \left[(j_y)_{i,j-1/2}+(j_y)_{i,j+1/2}\right]
    \times
    \left[(v_x)_{i-1/2,j}+(v_x)_{i+1/2,j}\right], \\
    (j_y v_y)_{i,j}
    &= \frac{1}{4}
    \left[(j_y)_{i,j-1/2}+(j_y)_{i,j+1/2}\right]
    \times
    \left[(v_y)_{i,j-1/2}+(v_y)_{i,j+1/2}\right].
\end{align}

The discrete differential operators are chosen to preserve the fluctuation-dissipation relaxation~\cite{BalboaUsabiaga2012-sh, Srivastava2023-nx}.
We use the following central differences.

\begin{description}
\item[Gradient] For a scalar field,
\begin{align}
    \big(\bm{\nabla}_x \rho\big)_{i+1/2, j} \ \to \frac{\rho_{i+1, j} - \rho_{i, j}}{\Delta x} .
\end{align}
\item[Divergence] For a vector field,
\begin{align}
    \big(\bm{\nabla} \cdot \bm{j}\big)_{i, j}
    &\to \frac{(j_x)_{i+1/2, j} - (j_x)_{i-1/2, j}}{\Delta x}
    + \frac{(j_y)_{i, j+1/2} - (j_y)_{i, j-1/2}}{\Delta y} .
\end{align}
For a tensor field, the divergence is applied row by row; for example,
\begin{align}
    \big(\bm{\nabla} \cdot \bm{\Pi}^{\mathrm{ran}}_{x}\big)_{i+1/2, j}
    &\to \frac{1}{\Delta x}
    \bigl[
    (\bm{\Pi}^{\mathrm{ran}}_{xx})_{i+1, j}
    -(\bm{\Pi}^{\mathrm{ran}}_{xx})_{i, j}
    \bigr] \nonumber \\
    &\quad + \frac{1}{\Delta y}
    \bigl[
    (\bm{\Pi}^{\mathrm{ran}}_{xy})_{i+1/2, j+1/2}
    -(\bm{\Pi}^{\mathrm{ran}}_{xy})_{i+1/2, j-1/2}
    \bigr].
\end{align}
\item[Laplacian] The Laplacian is evaluated at the same location as the original variable; for example,
\begin{align}
    \big(\bm{\nabla}^2 v_x\big)_{i+1/2,j}
    &\to \frac{(v_x)_{i+3/2,j}
    -2(v_x)_{i+1/2,j}
    +(v_x)_{i-1/2,j}}{(\Delta x)^2} \nonumber \\
    &\quad + \frac{(v_x)_{i+1/2,j+1}
    -2(v_x)_{i+1/2,j}
    +(v_x)_{i+1/2,j-1}}{(\Delta y)^2}.
\end{align}
\end{description}

\subsection{Boundary conditions}
\label{app2-3}
We impose periodic boundary conditions along $y$ and no-slip boundary conditions at $x=0,L$:
\begin{align}
    v_x(\bm{r}) &= 0 \quad {\rm at} \ \ x= 0, L, \\
    v_y(\bm{r}) &= \mp U/2 \quad {\rm at} \ \ x= 0, L, \\
    \partial_x \rho(\bm{r}) &= 0 \quad {\rm at} \ \ x= 0, L.
\end{align}

To implement the no-slip boundary conditions on the staggered grid, we add one ghost-cell row at each wall, extending the computational domain beyond the physical boundaries.
Figure~\ref{supfig1} illustrates this construction for $(L_x,L_y)=(4.0,4.0)$ and $\auv=1.0$, where a $4\times 5$ grid is used.
The regions $x\leq 0$ and $x\geq L$ represent the left and right walls, respectively.

For the left wall, the boundary conditions are discretized as
\begin{align}
    (v_x)_{-1/2, j} &= -(v_x)_{1/2, j}, \\
    (v_y)_{0, j+1/2} &= -U/2, \\
    (\partial_x \rho)_{0, j} &= 0.
\end{align}
In the implementation, the density condition is imposed through the equivalent relation
\begin{align}
    (j_x)_{-1/2,j} &= - (j_x)_{1/2,j}.
\end{align}
For the right wall, we use
\begin{align}
    (v_x)_{9/2, j} &= -(v_x)_{7/2, j}, \\
    (v_y)_{4, j+1/2} &= U/2, \\
    (j_x)_{9/2, j} &= -(j_x)_{7/2, j}.
\end{align}

\begin{table}[tb]
\centering
\begin{tabularx}{1\columnwidth}{@{\extracolsep{\fill}}c|c|c|c}
\hline
$\eta_0$ & Flow type & Relaxation steps & Samples \\
\hline
$0.100$ & Couette & $18,000,000$ & 3456 \\
$0.325$ & Couette & $8,000,000$ & 1152 \\
$0.470$ & Couette & $8,000,000$ & 2304 \\
$0.660$ & Couette & $8,000,000$ & 3456 \\
$0.325$ & Poiseuille & $20,000,000$ & 1152 \\
$0.470$ & Poiseuille & $20,000,000$ & 1152 \\
$0.660$ & Poiseuille & $20,000,000$ & 1152 \\
$0.325$ & Equilibrium & $4,000,000$ & 1152 \\
$0.470$ & Equilibrium & $4,000,000$ & 1152 \\
$0.660$ & Equilibrium & $4,000,000$ & 1152 \\
\hline
\end{tabularx}
\caption{
Relaxation steps and number of independent samples used in the fluctuating hydrodynamics simulations.
}
\label{suptab1}
\end{table}

\subsection{Averaging procedure}
\label{app2-5}

We use reduced units with $m=\sigma=T=1.0$ and a time step $dt=0.001$.
Unless otherwise stated, the simulations use $\rho_0=0.765$, $L=128$, $c_T^2=1000$, and $\zeta_0=1.0$.
The large value of $c_T^2$ keeps the fluid nearly incompressible.

Each run consists of a relaxation stage followed by an observation stage of $10,000,000$ steps, corresponding to $10,000$ time units.
Measurements are taken every 100 steps.
The reported averages are obtained by combining time averages over the observation stage with averages over independent noise realizations.
The relaxation time and number of samples are summarized in Table~\ref{suptab1}.

\begin{figure}[tb]
\begin{center}
\includegraphics[scale=1.0]{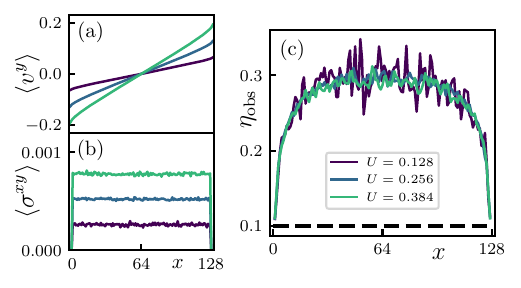}
\end{center}
\caption{
Wall-velocity dependence of fluid flows in the fluctuating hydrodynamics simulation.
(a) Velocity profiles, $\langle v^y(x) \rangle$.
(b) Shear stress profiles, $\langle \sigma^{xy}(x) \rangle$.
(c) Observed viscosity profiles, $\etaR(x)$.
All profiles are shown as functions of x.
The parameters are the same as in Fig.~\ref{fig1}, except that $L$ is fixed at $128$ and $U$ varies from $0.128$ (purple) to $0.384$ (green).
The plot of $\etaR(x)$ in (c) demonstrates that the observed viscosity does not depend on $U$, suggesting that our observations are within the linear response regime.
}
\label{supfig2}
\end{figure}
\section{Confirmation of the linear-response regime in fluctuating hydrodynamics simulation}
\label{app3}
In Fig.~\ref{fig1}, the fluctuating hydrodynamic simulations were performed at a fixed shear rate, $U/L=0.002$, for different system sizes.
Here we check that the observed system-size dependence of the shear stress and viscosity is not affected by this choice of shear rate.

To this end, we perform additional simulations at fixed $L=128$ while varying $U$.
As shown in Fig.~\ref{supfig2}, changing $U/L$ changes the velocity gradient and the shear stress [Figs.~\ref{supfig2}(a) and (b)].
The observed viscosity $\etaR(x)$, however, remains independent of $U/L$ [Fig.~\ref{supfig2}(c)].
This confirms that the simulations reported in Fig.~\ref{fig1} are in the linear-response regime.

\section{Derivation of Eqs.~(\ref{eq:velocity profile in pt})-(\ref{eq:etaR in pt})}
\label{app4}

This appendix derives the analytical expressions for the velocity profile and the local observed viscosity in the Couette geometry, Eqs.~(\ref{eq:velocity profile in pt}) and (\ref{eq:etaR in pt}).
The calculation is based on incompressible fluctuating hydrodynamics with no-slip boundary conditions.
We first reduce the equations to a tractable stochastic equation for the velocity field and then evaluate the fluctuation-induced correction perturbatively up to second order in the nonlinear coupling.

\subsection{Setup of analytic calculations}
\label{app4-1}

For the analytical treatment, we take the incompressible limit of the fluctuating hydrodynamic equations introduced in Sec.~\ref{sec2-1}.
In this limit, the density is fixed at $\rho_0$, and the velocity field obeys
\begin{align}
\pd{\bm{v}}{t} + \epsilon \bm{v} \cdot \nabla \bm{v} &= - \frac{1}{\rho_0} \nabla p + \nu_0 \Delta \bm{v} + \nabla \cdot \bm{\Pi}^{\mathrm{ran}}, \label{eq: incompressible Navier-Stokes equation} \\
\nabla \cdot \bm{v} &= 0 \label{eq: incompressible condition}
\end{align}
with
\begin{align}
    \mybra \bm{\Pi}^{\mathrm{ran}}_{ab}(\bm{r},t)
    \bm{\Pi}^{\mathrm{ran}}_{cd}(\bm{r}',t') \myket
    &= \frac{2 \nu_0 k_B T}{\rho_0}
    \delta(\bm{r}-\bm{r}')\delta(t-t') \biggl[
    \left(\delta_{ac}\delta_{bd}+\delta_{ad}\delta_{bc}\right)
    - \delta_{ab}\delta_{cd} \biggr].
    \label{eq: incompressible random stress tensor}
\end{align}
Here, $\nu_0 := \eta_0/\rho_0$ and $\epsilon$ is a perturbative parameter, which will eventually be set to $1$.
The pressure $p$ is an auxiliary field determined by the constraint
$\nabla\cdot\bm{v}=0$.
Taking the divergence of Eq.~(\ref{eq: incompressible Navier-Stokes equation}) gives
\begin{align}
    \frac{1}{\rho_0} \Delta p
    = -\epsilon (\partial_a v_b) (\partial_b v_a)
    + \partial_a \partial_b \bm{\Pi}^{\mathrm{ran}}_{ab}.
    \label{eq: incompressible pressure equation}
\end{align}
Substituting this expression for $p$ back into Eq.~(\ref{eq: incompressible Navier-Stokes equation}), we obtain
\begin{align}
    & \pd{\bm{v}}{t}
    + \epsilon \bm{v} \cdot \nabla \bm{v}
    + \epsilon \nabla \Delta^{-1}(\partial_a v_b)(\partial_b v_a)
    = \nu_0 \Delta \bm{v} + \bm{f},
    \label{eq: incompressible Navier-Stokes equation, mod} \\
    & \langle f^a(\bm{r},t) f^b(\bm{r}',t')\rangle
    = - \frac{2\nu_0 k_B T}{\rho_0}
    P^{ab}(\nabla)
    \delta(\bm{r}-\bm{r}') \delta(t-t'),
    \label{eq: incompressible random stress tensor, mod}
\end{align}
where $P^{ab}(\nabla):=\delta^{ab}\Delta-\partial_a\partial_b$.

To obtain analytic expressions, we make two simplifying approximations to Eq.~(\ref{eq: incompressible Navier-Stokes equation, mod}).
First, we neglect the nonlocal pressure-feedback term
$\nabla \Delta^{-1}(\partial_a v_b)(\partial_b v_a)$.
This term enforces incompressibility at the nonlinear level and can affect fluctuation correlations.
Second, we replace the projected noise correlation by its diagonal part, thereby neglecting the off-diagonal components of $P^{ab}(\nabla)$.
With these approximations, we use the simplified stochastic equation
\begin{align}
    \pd{\bm{v}}{t} + \epsilon \bm{v} \cdot \nabla \bm{v} &= \nu_0 \Delta \bm{v} + \bm{f}, \label{eq: iNSeq, to solve} \\
    \langle f^a(\bm{r},t) f^b(\bm{r}',t')\rangle &= - \frac{2\nu_0 k_B T}{\rho_0} \delta^{ab} \Delta \delta(\bm{r}-\bm{r}') \delta(t-t'). \label{eq: irst, to solve}
\end{align}
These approximations are introduced to make the calculation analytically tractable.
Their adequacy is assessed a posteriori by comparison with the full fluctuating hydrodynamic simulations.
As shown in Sec.~\ref{sec3}, the resulting expressions reproduce the spatial profiles obtained numerically, although although the prefactor of the fluctuation-induced correction is treated as a fitting parameter.

We solve this equation under no-slip boundary conditions at $x=0, L$:
\begin{equation}
    \bm{v} = \bm{0} \quad {\rm at} \ \ x=0,
    \qquad
    \bm{v} = U\bm{e}_y \quad {\rm at} \ \ x=L.
    \label{eq: BCapp}
\end{equation}
Periodic boundary conditions are imposed along the $y$ axis.
This setup corresponds to the Couette flow configuration considered in the main text.

\subsection{Decomposition into noise-averaged velocity and fluctuations}
\label{app4-2}

We decompose the velocity field into its noise average and fluctuation:
\begin{align}
   v^a(\bm{r},t) &= \mybra v^a(\bm{r},t) \myket + u^a(\bm{r},t).
\end{align}
Substitution into Eq.~(\ref{eq: iNSeq, to solve}) gives the equation for the noise-averaged velocity,
\begin{align}
    \frac{\partial\mybra \bm{v} \myket}{\partial t}
    + \epsilon \Big(
    \mybra \bm{v} \myket \cdot \nabla \mybra \bm{v} \myket
    + \mybra \bm{u} \cdot \nabla \bm{u} \myket\Big)
    &= \nu_0 \Delta \mybra \bm{v} \myket,
    \label{eq: iNSeq_det, to solve}
\end{align}
and the equation for the fluctuation,
\begin{align}
    \frac{\partial \bm{u}}{\partial t}
    + \epsilon \Big(
    \mybra \bm{v} \myket \cdot \nabla \bm{u}
    + \bm{u} \cdot \nabla \mybra \bm{v} \myket
    + \bm{u} \cdot \nabla \bm{u}
    - \mybra \bm{u} \cdot \nabla \bm{u} \myket\Big)
    &= \nu_0 \Delta \bm{u} + \bm{f}.
    \label{eq: iNSeq_flu, to solve}
\end{align}
The boundary conditions become
\begin{equation}
    \mybra \bm{v} \myket = \bm{0} \quad {\rm at} \ \ x=0,
    \qquad
    \mybra \bm{v} \myket = U\bm{e}_y \quad {\rm at} \ \ x=L,
    \label{eq: BCapp: det}
\end{equation}
and
\begin{align}
\bm{u} = \bm{0} \quad {\rm at} \ \ x=0,L .
\label{eq: BCapp: flu}
\end{align}

In the steady Couette state, the mean velocity has the form
\begin{align}
    \langle v^x \rangle = 0,
    \qquad
    \langle v^y \rangle \equiv V(x).
\end{align}
Then Eq.~(\ref{eq: iNSeq_det, to solve}) reduces to the stress-balance equation
\begin{align}
    \partial_b \left[\rho_0\epsilon \Big(\langle v^a \rangle \langle v^b \rangle + \langle u^a u^b \rangle\Big) - \eta_0 \partial_b \langle v^a \rangle \right] = 0,
\end{align}
or, for the $y$ component,
\begin{align}
    \rho_0\epsilon \langle u^x u^y \rangle - \eta_0 \partial_x \langle v^y \rangle = {\rm const} := - \langle \sigma^{xy}\rangle.
    \label{eq: stressbalance}
\end{align}
Here, $\eta_0=\rho_0\nu_0$, and $\langle \sigma^{xy}\rangle$ is the physical noise-averaged shear stress.

\subsection{Perturbation theory}
\label{app4-3}

We solve Eqs.~(\ref{eq: iNSeq_flu, to solve}) and
(\ref{eq: stressbalance}) perturbatively in $\epsilon$.
The mean velocity and fluctuation are expanded as
\begin{align}
\big\langle v^a(\bm{r},t) \big\rangle
&= \big\langle v^a_{(0)}(\bm{r},t) \big\rangle
 + \epsilon \big\langle v^a_{(1)}(\bm{r},t) \big\rangle
 + \cdots, \label{eq:expand_det}\\
u^a(\bm{r},t)
&= u^a_{(0)}(\bm{r},t)
 + \epsilon u^a_{(1)}(\bm{r},t) + \cdots. \label{eq:expand_flu}
\end{align}

We keep terms up to $O(\epsilon^2)$ and denote the truncated mean velocity by
\begin{align}
    \langle v^y_{(\leq 2)}\rangle := \langle v^y_{(0)}\rangle + \epsilon\langle v^y_{(1)}\rangle + \epsilon^2 \langle v^y_{(2)}\rangle.
    \label{eq:def_upto2}
\end{align}
Substitution into Eq.~(\ref{eq: stressbalance}) gives
\begin{align}
    \eta_0 \partial_x \langle v^y_{(\leq 2)}\rangle
    &= \langle \sigma^{xy}\rangle
    + \rho_0\epsilon \langle u^x_{(0)} u^y_{(0)} \rangle
    + \rho_0\epsilon^2
    \Big[
    \langle u^x_{(1)} u^y_{(0)} \rangle
    + \langle u^x_{(0)} u^y_{(1)} \rangle
    \Big].
    \label{eq:coue_uptosecond}
\end{align}
Thus, the mean profile is determined once the zeroth- and first-order fluctuations are known.
They obey
\begin{align}
    \frac{\partial u^a_{(0)}}{\partial t} = \nu_0 \Delta u^a_{(0)} + f^a,
    \label{eq:coue_flu_zero}
\end{align}
and
\begin{align}
    \frac{\partial u^a_{(1)}}{\partial t}
    &+ \Big[\langle v^y_{(0)}\rangle \partial_y u_{(0)}^a
    + u_{(0)}^b \partial_b \langle v^a_{(0)}\rangle + u_{(0)}^b \partial_b u_{(0)}^a
    - \langle u_{(0)}^b \partial_b u_{(0)}^a \rangle \Big]
    = \nu_0 \Delta u^a_{(1)}.
    \label{eq:coue_flu_first}
\end{align}

The boundary conditions are imposed order by order:
\begin{equation}
        \langle v^y_{(\leq 2)}\rangle = 0 \quad {\rm at} \ \ x=0,
        \qquad
        \langle v^y_{(\leq 2)}\rangle = U \quad {\rm at} \ \ x=L,
    \label{eq:coue_bou_second}
\end{equation}
\begin{equation}
        \bm{u}_{(0)} = \bm{0} \quad {\rm at} \ \ x=0,L,
        \qquad
        \bm{u}_{(1)} = \bm{0} \quad {\rm at} \ \ x=0,L.
    \label{eq:coue_flubou_first}
\end{equation}

\subsubsection*{Zeroth-order solution}
At zeroth order, Eqs.~(\ref{eq:coue_uptosecond}) and
(\ref{eq:coue_bou_second}) reduce to
\begin{align}
    \eta_0 \partial_x \langle v^y_{(0)}\rangle = \langle \sigma^{xy}\rangle,
    \label{eq:coue_zeroth}
\end{align}
\begin{equation}
        \langle v^y_{(0)}\rangle = 0 \quad {\rm at} \ \ x=0,
        \qquad
        \langle v^y_{(0)}\rangle = U \quad {\rm at} \ \ x=L.
\end{equation}
Since $\langle \sigma^{xy}\rangle$ is constant, Eq.~(\ref{eq:coue_zeroth}) gives
\begin{align}
    \langle v^y_{(0)}\rangle = \frac{\langle \sigma^{xy}\rangle}{\eta_0} x + C,
    \label{eq:coue_zero_int}
\end{align}
where $C$ is an integration constant.
The boundary conditions determine
\begin{align}
    \langle \sigma^{xy}\rangle &= \eta_0 \dot{\gamma}, \\
    C &= 0,
\end{align}
where $\dot{\gamma}:=U/L$.
Thus
\begin{align}
    \langle v^y_{(0)}\rangle = \dot{\gamma} x.
    \label{eq:coue_zero}
\end{align}

\subsubsection*{Second-order solution}
The first-order correction to the mean profile vanishes, as shown below from
$\langle u^x_{(0)}u^y_{(0)}\rangle=0$.
We therefore proceed to the leading nonzero correction, which is of order
$O(\epsilon^2)$.

Substituting Eq.~(\ref{eq:coue_zero}) into Eq.~(\ref{eq:coue_flu_first}), we obtain
\begin{align}
    \frac{\partial u^a_{(1)}}{\partial t}
    &+ \Big[\dot{\gamma} x \partial_y u_{(0)}^a
    + \dot{\gamma} u_{(0)}^x \delta_{ay} + u_{(0)}^b \partial_b u_{(0)}^a
    - \langle u_{(0)}^b \partial_b u_{(0)}^a \rangle \Big]
    = \nu_0 \Delta u^a_{(1)}.
    \label{eq:coue_flu_first_mod}
\end{align}

To evaluate the fluctuation correlations in Eq.~(\ref{eq:coue_uptosecond}), we use the Green function
$G(\bm{r},\bm{r}',\tau)$ defined by
\begin{align}
    \Big(\frac{\partial}{\partial \tau} - \nu_0 \Delta \Big) G(\bm{r}, \bm{r}', \tau) = \delta(\bm{r}-\bm{r}') \delta(\tau),
\end{align}
with the Dirichlet boundary condition
\begin{align}
    G(\bm{r}, \bm{r}', \tau) = 0 \quad {\rm at} \ \ x=0,L \ {\rm or} \ x'=0,L.
\end{align}
Its explicit form is
\begin{align}
      G(\bm{r},\bm{r}',\tau)
      &= \theta(\tau)\int_k
      e^{ik_y (y-y')}
      \sin(k_x x)\sin(k_x x')
      \times
      e^{-\nu_0 (k_x^2+k_y^2)\tau},
      \label{eq:Green_expli}
\end{align}
where $k_x:=\pi n_x/L$ and
\begin{align}
    \int_k:=\frac{2}{L}\sum_{n_x=1}^{L/\auv} \int^\infty_{-\infty} \frac{d k_y}{2\pi}.
\end{align}
The form of this solution reflects the translational symmetry along the $y$ axis (plane wave expansion) and the confinement of the fluid at $x=0, L$ (sine function expansion).

Using this Green function, the zeroth- and first-order fluctuation fields,
$u^a_{(0)}$ and $u^a_{(1)}$, are represented as
\begin{align}
    \uz^a(\bm{r},t)
    &= \int G(\bm{r},\bm{r}',\tau)f^a(\bm{r}',t-\tau),
    \label{eq:solzero}\\
    \uf^a(\bm{r},t)
    &= -\int G(\bm{r},\bm{r}',\tau) \biggl[\dot{\gamma} x' \partial_{y'} u_{(0)}^a + \dot{\gamma} u_{(0)}^x \delta_{ay} + u_{(0)}^b \partial_b u_{(0)}^a - \langle u_{(0)}^b \partial_b u_{(0)}^a \rangle\biggr]\biggr\vert_{\bm{r}',t-\tau},\label{eq:solfirst}
\end{align}
where we have introduced the shorthand notation:
\begin{align}
    \int = \int^\infty_0 d\tau\int^L_{0}dx'\int^\infty_{-\infty} dy'.
\end{align}
Because $u_{(0)}^a$ is a Gaussian field generated by Eq.~(\ref{eq:coue_flu_zero}),
\begin{align}
    \langle  u_{(0)}^x(\bm{r},0) u_{(0)}^y (\bm{r},0)\rangle = 0, \label{eq:u0xu0y}\\
    \langle  u_{(0)}^a(\bm{r},0) u_{(0)}^b(\bm{r},0) u_{(0)}^c (\bm{r},0)\rangle = 0.
\end{align}
The first identity explains why the $O(\epsilon)$ term in Eq.~(\ref{eq:coue_uptosecond}) vanishes.
Together with the order-by-order boundary conditions, it gives $\langle v^y_{(1)}\rangle=0$.

We next evaluate the $O(\epsilon^2)$ correlations in Eq.~(\ref{eq:coue_uptosecond}).
Substituting Eq.~(\ref{eq:solfirst}) and using the vanishing of the three-point function, we obtain
\begin{align}
    \langle \uf^a(\bm{r},0)\uz^b(\bm{r},0)\rangle
    &= -\int G(\bm{r},\bm{r}',\tau)
    \biggl[
    \dot{\gamma} x'
    \langle \uz^b(\bm{r},0)
    \partial_{y'} \uz^a(\bm{r}',-\tau)\rangle \nonumber \\
    &\hspace{3cm} + \dot{\gamma} \delta_{ay}
    \langle \uz^b(\bm{r},0)
    \uz^x(\bm{r}',-\tau) \rangle
    \biggr],
\end{align}
and the components needed below are
\begin{align}
    \langle \uf^x(\bm{r},0)\uz^y(\bm{r},0)\rangle
    &= 0, \label{eq:u1xu0y} \\
    \langle \uf^y(\bm{r},0)\uz^x(\bm{r},0)\rangle
    &= - \dot{\gamma}
    \int G(\bm{r},\bm{r}',\tau)
    \langle \uz^x(\bm{r},0)
    \uz^x(\bm{r}',-\tau) \rangle.
    \label{eq:u0xu1y}
\end{align}

The noise correlation can be written as
\begin{align}
    \langle f^a(\bm{r},t)f^b(\bm{r}',t')\rangle
    &=\frac{2\nu_0 k_B T}{\rho_0}\delta^{ab}
    \int_k k^2 e^{ik_y (y-y')}
    \sin(k_x x)\sin(k_x x') \times \delta(t-t'),
\end{align}
which gives
\begin{align}
    \langle\uz^x(\bm{r},0)\uz^x(\bm{r}',-\tau) \rangle
    &=\frac{ k_B T}{\rho_0} \int_k e^{ik_y(y-y')}
    \sin(k_x x)\sin(k_x x') e^{-\nu_0 k^2\tau}.
\end{align}
\par\noindent Therefore,
\begin{align}
   \langle \uf^y(\bm{r},0)\uz^x(\bm{r},0)\rangle
   &= - \dot{\gamma} \int G(\bm{r},\bm{r}',\tau)
   \langle\uz^x(\bm{r},0)\uz^x(\bm{r}',-\tau) \rangle \nonumber \\
   &= - \frac{\dot{\gamma} k_B T}{2\nu_0 \rho_0 L}
   \sum_{k_x}\frac{1}{k_x}\sin^2(k_xx).
   \label{eq:u0xu1y_mod}
\end{align}

Substitution of Eqs.~(\ref{eq:u0xu0y}), (\ref{eq:u1xu0y}), and
(\ref{eq:u0xu1y_mod}) into Eq.~(\ref{eq:coue_uptosecond}) yields
\begin{align}
    \eta_0 \partial_x \langle v^y_{(\leq 2)}(\bm{r})\rangle
    &= \langle \sigma^{xy} \rangle
    - \epsilon^2 \frac{\rho_0 \dot{\gamma} k_B T}{2\eta_0 L}
    \sum_{k_x}\frac{1}{k_x}\sin^2(k_xx).
\end{align}
\par\noindent Thus the leading fluctuation correction is $O(\epsilon^2)$.

Integrating with respect to $x$ gives
\begin{align}
    \langle v^y_{(\leq 2)}(\bm{r})\rangle
    &= \frac{\langle \sigma^{xy} \rangle}{\eta_0} x
    - \epsilon^2 \frac{\rho_0 \dot{\gamma} k_B T}{4\eta_0^2 L}
    \times
    \sum_{k_x}\frac{1}{k_x}
    \left(x - \frac{\sin(2k_x x)}{2k_x} \right).
    \label{eq:coue_uptosecond_int}
\end{align}
The boundary conditions in Eq.~(\ref{eq:coue_bou_second}) determine the constant stress as
\begin{align}
    \langle \sigma^{xy} \rangle
    &= \eta_0 \dot{\gamma}
    \left[
    1 + \epsilon^2
    \frac{\rho_0 k_B T}{4\eta_0^2 L}
    \sum_{k_x}\frac{1}{k_x}
    \right].
\end{align}
Substituting this result back into Eq.~(\ref{eq:coue_uptosecond_int}), we obtain
\begin{align}
    \langle v^y_{(\leq 2)}(\bm{r})\rangle
    &= \dot{\gamma} x
    + \epsilon^2 \frac{\rho_0 \dot{\gamma} k_B T}{4\eta_0^2 L}
    \sum_{k_x}\frac{1}{k_x}\frac{\sin(2k_x x)}{2k_x}.
\end{align}
Finally, using $\etaR(x):=\langle \sigma^{xy} \rangle/\partial_x\langle v^y\rangle$, we find, to the same order,
\begin{align}
    \etaR(x)
    &=\eta_0 \biggl[
    1+\epsilon^2
    \frac{\rho_0 k_B T}{2\eta_0^2 L}
    \sum_{k_x}\frac{1}{k_x}\sin^2(k_xx)
    \biggr]
    +O(\epsilon^4).
\end{align}
Setting $\epsilon=1$, these expressions reproduce Eqs.~(\ref{eq:velocity profile in pt})-(\ref{eq:etaR in pt}) with the definition of $A$ in Eq.~(\ref{eq:numerical factor in pt}).

\section{Setup for MD simulations}
\label{app5}

This appendix summarizes the MD simulation setup, including the wall models, coarse-grained observables, and averaging procedure.
Although the fluid model is introduced in Sec.~\ref{sec2-1}, we briefly recall the basic setup here for completeness.

\subsection{Hamiltonian and equations of motion}

We consider $N$ particles of mass $m$ in a square domain $[0,L]\times[0,L]$.
We place walls at $x=0$ and $x=L$ and impose periodic boundary conditions along the $y$ axis.
Let $(\bm{r}_i,\bm{p}_i)$ denote the fluid variables.
We denote the left and right wall variables by $(\bm{r}_{{\rm left}, i}, \bm{p}_{{\rm left}, i})$ and $(\bm{r}_{{\rm right}, i}, \bm{p}_{{\rm right}, i})$.

We decompose the Hamiltonian as
\begin{align}
    H &= H_{\rm fluid} + H_{\rm left} + H_{\rm right}.
\end{align}
The fluid particles evolve according to Hamilton's equations of motion:
\begin{align}
    \frac{d \bm{r}_i}{dt} &= \frac{\bm{p}_i}{m}, \\
    \frac{d \bm{p}_i}{dt} &= -\frac{\partial H_{\rm fluid}}{\partial \bm{r}_i} -\frac{\partial H_{\rm left}}{\partial \bm{r}_i} -\frac{\partial H_{\rm right}}{\partial \bm{r}_i} + \bm{g}.
\end{align}
For Poiseuille flow, we add an external force $\bm{g}=g\bm{e}_y$.
We use three wall models: thermal hydrophobic, thermal hydrophilic, and frozen hydrophobic walls.
The main text focuses on the thermal hydrophobic wall, while Appendix~\ref{app6} uses the other two to test robustness.
Below we describe the left wall at $x=0$; we construct the right wall in the same way.

\subsubsection{Thermal hydrophobic wall}
The thermal hydrophobic wall consists of $N_{\rm left}$ particles arranged in three square-lattice layers, with the first layer at $x=-\sigma$ (see Fig.~\ref{supfig1}).
For the left wall, we use
\begin{align}
    H_{\rm left}
    &= \sum_{i=1}^{N_{\rm left}}
    \frac{\bm{p}_{{\rm left}, i}^2}{2m}
    + \sum_{i=1}^{N_{\rm left}}
    V^{\rm onsite}(\bm{r}_{{\rm left}, i}, t)
    + \sum_{i=1}^{N_{\rm left}} \sum_{j=1}^{N}
    V^{\rm wall}(|\bm{r}_{{\rm left}, i} -\bm{r}_j|).
\end{align}
The on-site potential takes the form
\begin{align}
    V^{\rm onsite}(\bm{r}, t)
    &= V_0 \Biggl[
    \cos\biggl(\frac{2\pi (x+\sigma/2)}{\sigma}\biggr)
    + \cos\biggl(\frac{2\pi (y\mp v_0t/2)}{\sigma}\biggr)
    \Biggr].
\end{align}
Here, $V_0$ is the strength of the on-site potential.
The time-dependent term moves the wall in the $y$ direction with velocity $\pm v_0$, thereby imposing Couette shear.
For the wall-fluid interaction, we use the Weeks--Chandler--Andersen (WCA) potential,
\begin{align}
    V^{\rm wall}(r) =
    \begin{cases}
        4\epsilon \Big\{\big(\frac{\sigma}{r}\big)^{12} - \big(\frac{\sigma}{r}\big)^{6} + \frac{1}{4} \Big\} & {\rm for} \ r<2^{1/6}\sigma \\
        0 & {\rm otherwise}.
    \end{cases}
    \label{eq:micro_hydrophobic_interaction}
\end{align}

The wall particles obey the Langevin equations
\begin{align}
    \frac{d \bm{r}_{{\rm left}, i}}{dt} &= \frac{\bm{p}_{{\rm left}, i}}{m}, \\
    \frac{d \bm{p}_{{\rm left}, i}}{dt} &= - \frac{\partial H}{\partial \bm{r}_{{\rm left}, i}} - \gamma \bm{p}_{{\rm left}, i} + \sqrt{2 \gamma T}\bm{\xi}_i(t),
    \label{eq:micro_thermal_wall_langevin}
\end{align}
where $\gamma$ represents the friction coefficient, and $\bm{\xi}_i(t)$ denotes Gaussian white noise with zero mean and unit variance.

\subsubsection{Thermal hydrophilic wall}
The thermal hydrophilic wall has the same particle arrangement and Langevin dynamics [Eq.~(\ref{eq:micro_thermal_wall_langevin})] as the thermal hydrophobic wall.
Only the wall-fluid interaction is changed from the WCA potential to the attractive Lennard--Jones (LJ) potential,
\begin{align}
    V^{\rm wall}(r) =
    \begin{cases}
        4\epsilon \Big\{\big(\frac{\sigma}{r}\big)^{12} - \big(\frac{\sigma}{r}\big)^{6} \Big\} & {\rm for} \ r<2.5\sigma \\
        0 & {\rm otherwise}.
    \end{cases}
    \label{eq:micro_hydrophilic_interaction}
\end{align}

\subsubsection{Frozen hydrophobic wall}
For the frozen hydrophobic wall, the wall particles are not evolved dynamically.
Instead, we impose their positions as
\begin{align}
    \bm{r}_{{\rm left}, i} = \biggl(\frac{i_x}{2}\sigma, \frac{i_y}{2}\sigma \pm \frac{v_0t}{2} \biggr)
\end{align}
with $i_x = -1,-2,-3$ and $i_y = 1,2,\cdots,2L/\sigma$.
Because these particles have no dynamical momenta, the corresponding wall Hamiltonian reduces to
\begin{align}
    H_{\rm left} = \sum_{i=1}^{N_{\rm left}} \sum_{j=1}^{N} V^{\rm wall}(|\bm{r}_{{\rm left}, i} -\bm{r}_j|).
\end{align}
We again use the WCA potential in Eq.~(\ref{eq:micro_hydrophobic_interaction}) for $V^{\rm wall}(r)$.

\subsection{Coarse-grained observables}
We compute hydrodynamic fields by coarse-graining microscopic particle configurations.
Let $\hat{\rho}(\bm{r};\Gamma_t)$, $\hat{j}^a(\bm{r};\Gamma_t)$, and $\hat{\Pi}^{ab}(\bm{r};\Gamma_t)$ denote the microscopic mass density, momentum density, and momentum current density for $\Gamma_t:=\{(\bm{r}_i(t),\bm{p}_i(t))\}_{i=1}^N$.
The explicit expressions for these quantities can be found in Refs.~\cite{Das2011-ao, Nakano2019-po}.

To obtain the one-dimensional steady-state profiles used in the main text, we first divide the system into bins of width $\auv$ along the $x$ axis and average the microscopic fields within each bin.
For example, for $x_i=i\auv$ ($i=0,1,\cdots,L/\auv$), the instantaneous coarse-grained density is
\begin{align}
    \rho(x_i;t)
    &:= \frac{1}{\auv} \int_{x_i -\auv /2}^{x_i + \auv /2} dx \frac{1}{L} \int_{0}^{L} dy\, \hat{\rho}(\bm{r};\Gamma_t).
\end{align}
The steady density profile is then obtained by time averaging over the observation interval,
\begin{align}
    \langle \rho(x_i)\rangle
    := \frac{1}{\tau_{\rm obs}}\int_0^{\tau_{\rm obs}} dt\,\rho(x_i;t).
\end{align}
The momentum density $\langle\bm{j}(x_i)\rangle$ and momentum current $\langle\bm{\Pi}_{ab}(x_i)\rangle$ are calculated in the same way.

Using these time-averaged fields, we compute the velocity field and stress tensor as
\begin{align}
    \langle \bm{v}(x_i) \rangle&:= \frac{\langle \bm{j}(x_i) \rangle}{\langle \rho(x_i) \rangle}, \\
    \langle \bm{\sigma}_{ab}(x_i) \rangle&:= -\langle \bm{\Pi}_{ab}(x_i) \rangle + \langle j_a(x_i) \rangle \langle v_b(x_i) \rangle.
\end{align}

Figure~\ref{supfig1} shows the correspondence between the MD observables and the grid used in fluctuating hydrodynamics.
In the comparison shown in the main text, we set $\auv=\sigma$.
The velocity field is defined at $x=0,\sigma,\cdots,L$ in both descriptions.
The shear stress is evaluated at different grid points in the two descriptions; because the steady-state shear stress is spatially homogeneous, we use its spatial average.
The observed viscosity is then evaluated as $\etaR(x)=\sigma^{xy}(x)/\partial_x v^y(x)$.

For the equilibrium momentum correlation $C_{\rm eq}(t)$ defined in Eq.~(\ref{eq: definition of time correlation function of the momentum field}), we use two-dimensional bins of width $\auv$ and calculate
\begin{align}
C_{\rm eq}(t) &= \frac{1}{\tau_{\rm obs}}\int_0^{\tau_{\rm obs}} ds \frac{1}{2N_{\mathcal{B}}} \sum_{{ij}\in \mathcal{B}} \bm{j}(\bm{r}_{ij};t+s) \cdot \bm{j}(\bm{r}_{ij};s)
\end{align}
with
\begin{align}
    \bm{j}(\bm{r}_i;t) := \frac{1}{\auv^2} \int_{x_i -\auv /2}^{x_i + \auv /2} dx \int_{y_i -\auv /2}^{y_i + \auv /2} dy \hat{\bm{j}}(\bm{r};\Gamma_t).
\end{align}
Here, $\bm{r}_{ij}=(i\auv, j\auv)$, and we restrict the sum to the central region $\mathcal{B}:= [L/4, 3L/4] \times [L/4, 3L/4]$.
$N_{\mathcal{B}}$ denotes the number of bins in $\mathcal{B}$, and the time average runs over $\tau_{\rm obs}$ in equilibrium.

\begin{table}[tb]
\centering
\begin{tabularx}{1\columnwidth}{@{\extracolsep{\fill}}c|c|c|c}
\hline
Potential & Flow type & Relaxation steps & Samples \\
\hline
$V(\delta)=10\delta^2$ & Couette & $18,000,000$ & 3456 \\
$V(\delta)=10\delta^4$ & Couette & $8,000,000$ & 1152 \\
$V(\delta)=10\delta^6$ & Couette & $8,000,000$ & 2304 \\
$V(\delta)=10\delta^2$ & Poiseuille & $18,000,000$ & 3456 \\
$V(\delta)=10\delta^4$ & Poiseuille & $8,000,000$ & 1152 \\
$V(\delta)=10\delta^6$ & Poiseuille & $8,000,000$ & 2304 \\
$V(\delta)=10\delta^2$ & Equilibrium & $18,000,000$ & 3456 \\
$V(\delta)=10\delta^4$ & Equilibrium & $8,000,000$ & 1152 \\
$V(\delta)=10\delta^6$ & Equilibrium & $8,000,000$ & 2304 \\
\hline
\end{tabularx}
\caption{
Relaxation steps and number of independent samples used for averaging in the MD simulations.
}
\label{suptab2}
\end{table}

\subsection{Averaging procedure}
We use reduced units with $m=\sigma=T=1.0$ and fix $V_0=50.0$, $\epsilon=1.0$, and $\gamma=1.0$.
We perform all simulations with LAMMPS using the velocity-Verlet algorithm with time step $dt=0.002$.

We divide each simulation into a relaxation phase and an observation phase.
Starting from a random particle configuration with Maxwell-Boltzmann velocities at temperature $T$, we evolve the system until it reaches the target steady or equilibrium state.
During the observation phase, we collect measurements every $100$ steps, corresponding to $100dt=0.2$.
The results presented in the main text combine the time averages defined above with averages over independent runs with different initial configurations and, when a stochastic thermostat is used, different noise realizations.
The relaxation steps and the number of independent samples are summarized in Table~\ref{suptab2}.

\clearpage
\section{Supplementary analyses of MD simulations}
\label{app6}

This appendix presents supplementary MD analyses that support the results of the main text.
We examine the dependence on microscopic wall properties, the range of linear response, and the system-size dependence of the observed viscosity.

\subsection{Effects of microscopic wall properties}
The main text uses thermal hydrophobic walls.
To test the sensitivity to the microscopic wall model, we compare this setup with frozen hydrophobic walls and thermal hydrophilic walls in the same Couette geometry as in Fig.~\ref{fig2}.

\begin{figure}[ht]
\begin{center}
\includegraphics[scale=1.0]{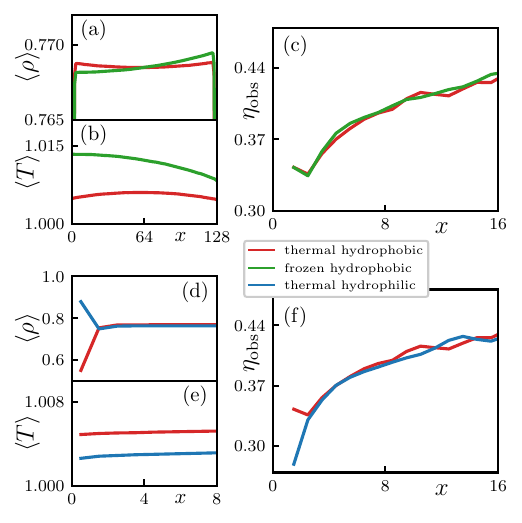}
\end{center}
\caption{
Effects of microscopic wall properties on fluid flows in the MD simulations.
(a-c) Comparison of thermal hydrophobic walls (red) and frozen hydrophobic walls (green): (a) density profiles, (b) temperature profiles, and (c) observed viscosity profiles.
(d-f) Comparison of thermal hydrophobic walls (red) and thermal hydrophilic walls (cyan), focusing on the near-wall region: (d) density profiles, (e) temperature profiles, and (f) observed viscosity profiles.
The interparticle potential is fixed at $V(\delta) = 10 \delta^2$, and the parameters are fixed at $\rho_0=0.765$, $T=1.0$, $U/L=0.002$, and $L=128.0$.
}
\label{supfig3}
\end{figure}

Figures~\ref{supfig3}(a)--(c) compare thermal and frozen hydrophobic walls.
In both simulations, the wall at $x=128$ is thermal and hydrophobic, whereas the wall at $x=0$ is either thermal hydrophobic or frozen hydrophobic.
The thermal wall exchanges energy with the fluid, while the frozen wall does not.
This difference produces a slightly higher temperature near the frozen wall [Fig.~\ref{supfig3}(b)], but the observed viscosity profiles remain quantitatively similar [Fig.~\ref{supfig3}(c)].
Thus, the near-wall reduction of $\etaR(x)$ is not controlled by direct thermalization at the wall.

Figures~\ref{supfig3}(d)--(f) compare thermal hydrophobic and thermal hydrophilic walls.
The wall-fluid interaction strongly affects the density in the first layer: the density decreases near the hydrophobic wall and increases near the hydrophilic wall [Fig.~\ref{supfig3}(d)].
Nevertheless, the viscosity profiles beyond the first layer are nearly identical [Fig.~\ref{supfig3}(f)].

Taken together, these comparisons show that the observed viscosity profile is robust against changes in the microscopic wall model.
This robustness supports the hydrodynamic description used in the main text, in which the wall is modeled by a no-slip boundary condition.

\subsection{Confirmation of the linear-response regime}
The main text uses a fixed shear rate $U/L=0.002$.
Here we confirm that this value lies in the linear-response regime.

\begin{figure}[ht]
\begin{center}
\includegraphics[scale=1.0]{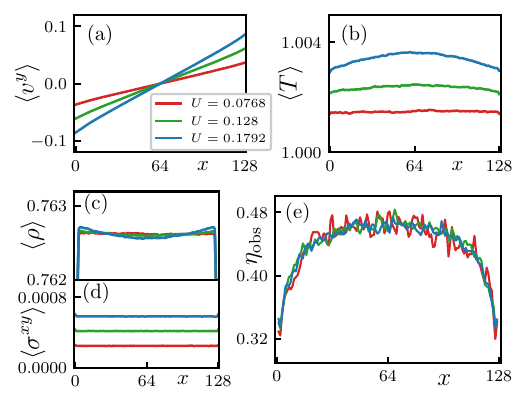}
\end{center}
\caption{
Wall-velocity dependence of fluid flows in the MD simulations.
(a) Velocity profiles, $\langle v^y(x) \rangle$.
(b) Temperature profiles, $\langle T(x) \rangle$.
(c) Density profiles, $\langle \rho(x) \rangle$.
(d) Shear stress profiles, $\langle \sigma^{xy}(x) \rangle$.
(e) Observed viscosity profiles, $\etaR(x)$.
All profiles are shown as functions of $x$.
The parameters are the same as those of the thermal hydrophobic wall in Fig.~\ref{supfig3}, except that $L$ is fixed at $128$ and $U$ varies from $0.0768$ (red) to $0.1792$ (cyan).
The plot of $\etaR(x)$ in (e) demonstrates that the observed viscosity does not depend on $U/L$, suggesting that our observations are within the linear response regime.
}
\label{supfig4}
\end{figure}

Figure~\ref{supfig4} shows MD results for several values of $U/L$ with thermal hydrophobic walls at $x=0$ and $x=128$.
As $U/L$ increases, the shear stress increases accordingly [Figs.~\ref{supfig4}(a) and (d)].
In contrast, the observed viscosity profile $\etaR(x)$ collapses onto a single curve [Fig.~\ref{supfig4}(e)].
This shear-rate independence confirms that the simulations used in the main text are in the linear-response regime.
We also note that the bulk temperature shows only a slight increase [Fig.~\ref{supfig4}(b)], because the fluid is thermalized only through the walls.
This temperature variation is small and does not affect the viscosity profile.

\subsection{Confirmation of system-size dependence}

The fluctuating-hydrodynamics simulations in Fig.~\ref{fig1} show two distinct trends: near the walls, $\etaR(x)$ is insensitive to $L$ and approaches the bare viscosity, whereas in the bulk it grows logarithmically with $L$.
We test whether the same behavior appears in MD simulations by varying the system size.

\begin{figure}[ht]
\begin{center}
\includegraphics[width=\textwidth]{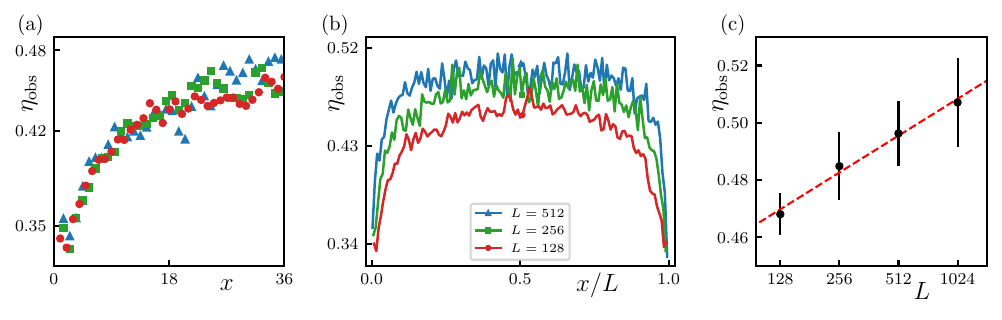}
\end{center}
\caption{
System-size dependence of observed viscosity in the MD simulations.
(a) Near-wall behavior of the observed viscosity $\etaR(x)$, extracted from (b).
(b) Overall behavior of $\etaR(x)$ as a function of the scaled position $x/L$.
(c) System-size dependence of $\etaR$ in the bulk region [$0.4 \leq x/L \leq 0.6$], extracted from (b).
The red line in (c) indicates the logarithmic divergence of $\etaR$.
The parameters are the same as in Fig.~\ref{supfig4}, but $U/L$ is fixed at $0.0010$ and $L$ varies from $128$ (red) to $512$ (blue).
}
\label{supfig5}
\end{figure}

Figure~\ref{supfig5}(b) shows that the bulk value of $\etaR(x)$ increases with $L$.
The bulk average, plotted in Fig.~\ref{supfig5}(c), follows a logarithmic dependence on $L$, consistent with anomalous transport in two-dimensional fluids and with the fluctuating-hydrodynamics result in Fig.~\ref{fig1}(d).

By contrast, Fig.~\ref{supfig5}(a) shows that the near-wall profile of $\etaR(x)$, plotted against the unscaled coordinate $x$, is almost independent of $L$.
Thus, the MD simulations reproduce both the bulk system-size dependence and the near-wall suppression of this dependence.

\section{Validation for other interparticle potentials}
\label{app7}

In the main text, the proposed protocols are tested for the atomic system with $V(\delta)=10\delta^2$ (Fig.~\ref{fig2}).
Here we repeat the same analysis for $V(\delta)=10\delta^4$ and $V(\delta)=10\delta^6$.
This comparison tests whether the estimated bare viscosity and the fluctuating-hydrodynamic description are robust against changes in the microscopic interaction.

\begin{figure}[tb]
\begin{center}
\includegraphics[scale=1.0]{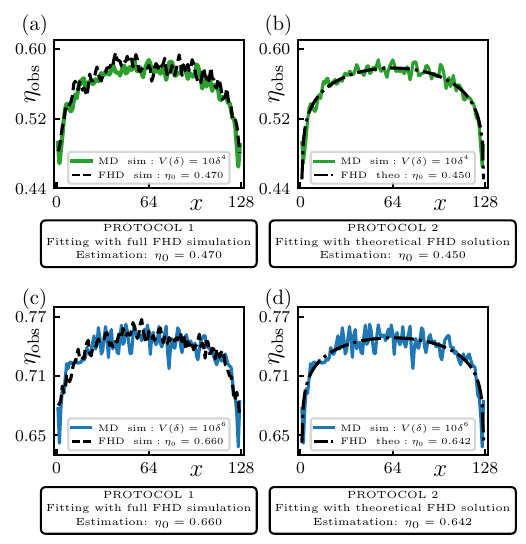}
\end{center}
\caption{
Validation of the proposed protocols for other interparticle potentials.
The same Couette analysis as in Figs.~\ref{fig2}(b) and (c) is applied to atomic systems with $V(\delta)=10\delta^4$ and $V(\delta)=10\delta^6$.
The parameters are fixed at $\rho_0=0.765$, $T=1.0$, $U/L=0.0014$, and $L=128$.
(a, b) Results for $V(\delta) = 10 \delta^4$.
(c, d) Results for $V(\delta) = 10 \delta^6$.
}
\label{supfig6}
\end{figure}

Figure~\ref{supfig6} shows Protocols 1 and 2 applied to these two systems in the same Couette setup as in Figs.~\ref{fig2}(b) and (c).
For both interparticle potentials, fluctuating hydrodynamics with the estimated $\eta_0$ reproduces the MD results quantitatively.
The obtained bare viscosity values are summarized in Table~\ref{tab1} in the main text.

\begin{figure}[tb]
\begin{center}
\includegraphics[scale=1.0]{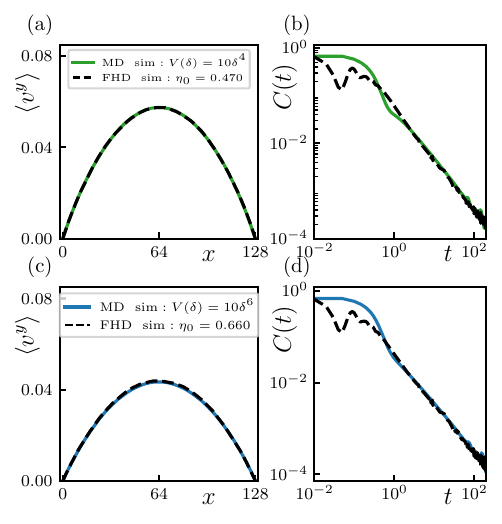}
\end{center}
\caption{
Independent validation of the bare viscosity estimated for other interparticle potentials.
Fluctuating hydrodynamics uses the value of $\eta_0$ estimated from Protocol 1 in Fig.~\ref{supfig6}.
(a, b) Results for $V(\delta) = 10 \delta^4$: (a) Velocity field for Poiseuille flow and (b) Time correlation function in equilibrium.
(c, d) Results for $V(\delta) = 10 \delta^6$: (c) Velocity field for Poiseuille flow and (d) Time correlation function in equilibrium.
The setup and parameters are the same as in Figs.~\ref{fig2}(d) and (e).
}
\label{supfig7}
\end{figure}

As an independent validation, we use the $\eta_0$ estimated from Protocol 1 to predict observables in the Poiseuille and equilibrium setups.
Figure~\ref{supfig7} compares the Poiseuille velocity profile and the equilibrium momentum-density correlation function $C_{\rm eq}(t)$ with fluctuating-hydrodynamic results.
The agreement in both cases shows that the same bare viscosity describes not only the Couette measurement used for the estimate but also different flow and equilibrium observables.

These results support the applicability of the proposed protocols beyond a specific choice of interparticle potential.
They also show that fluctuating hydrodynamics remains quantitative at the microscopic length and time scales resolved in the MD simulations.


\section{Bare-viscosity estimation in Poiseuille geometry}
\label{app8}

In the main text, we estimated the bare viscosity from the near-wall viscosity profile in Couette flow.
This appendix applies the same idea to Poiseuille flow.
We first derive the perturbative analytical expression for the noise-averaged velocity profile in Poiseuille flow and define the corresponding local observed viscosity.
We then use this expression to formulate a Poiseuille version of Protocol 2 and compare it with MD simulations.

\subsection{Measurement protocol in Poiseuille flow}

Consider a fluid confined between two fixed parallel walls located at $x=0$ and $x=L$.
A steady flow is driven by a constant external force acting on the fluid in the $y$ direction.
Within the perturbative fluctuating-hydrodynamic analysis used in Appendix~\ref{app4}, the velocity field up to second order in the nonlinear term is
\begin{align}
    \langle v^y_{(\leq 2)}(\bm{r})\rangle
    &= \frac{\rho_0 g}{2\eta_0}x(L-x)
    + g \frac{A_{\rm P}}{L} \sum_{k_x, k'_x}
    \frac{a(k_x, k'_x)}{\sqrt{k_x^2 + (k'_x)^2}} \nonumber \\
    &\qquad {}\left[
    \frac{\sin((k_x-k'_x)x)}{k_x-k'_x}
    - \frac{\sin((k_x+k'_x)x)}{k_x+k'_x}
    \right]
    \label{eq:pois_uptosecond_main}
\end{align}
with
\begin{align}
    A_{\rm P} := \frac{\sqrt{2}\rho_0^3 k_B T}{8\eta_0^3},
\end{align}
where $g$ represents the constant external force per unit mass.
The derivation is given in the next subsection.

For deterministic Poiseuille flow with viscosity $\eta$, the local curvature satisfies
\begin{align}
    \eta\, \partial_x^2 \langle v^y(x)\rangle = -\rho_0 g.
\end{align}
We therefore define the local observed viscosity in the Poiseuille geometry by
\begin{align}
    \etaR^{\rm poi}(x)
    := -\frac{\rho_0 g}{\partial_x^2 \langle v^y(x)\rangle}.
\end{align}

Using Eq.~(\ref{eq:pois_uptosecond_main}), we obtain
\begin{align}
    \etaR^{\rm poi}(x)
    &= \eta_0 \Biggl[1 + 2\frac{A_{\rm P}}{L} \sum_{k_x, k'_x} \frac{a(k_x, k'_x)}{\sqrt{k_x^2 + (k'_x)^2}}
    \Big\{\sin[(k_x-k'_x)x] - \sin[(k_x+k'_x)x]\Big\} \Biggr]^{-1} \nonumber \\
    &= \eta_0 \Biggl[1 - 2\frac{A_{\rm P}}{L} \sum_{k_x, k'_x} \frac{a(k_x, k'_x)}{\sqrt{k_x^2 + (k'_x)^2}} \nonumber \\
    &\qquad \Big\{\sin[(k_x-k'_x)x] - \sin[(k_x+k'_x)x]\Big\} \Biggr] + O(A_{\rm P}^2).
\end{align}
This expression is the Poiseuille counterpart of the Couette result in Eq.~(\ref{eq:etaR in pt}).
In both geometries, thermal fluctuations produce a spatially dependent correction to the bare viscosity.
At the walls, the correction vanishes:
\begin{align}
    \etaR^{\rm poi}(x=0) = \etaR^{\rm poi}(x=L) = \eta_0.
\end{align}
Thus, the bare viscosity governs the near-wall behavior also in Poiseuille flow.

This observation leads to the following Poiseuille version of Protocol 2.

\textbf{Protocol 2 (Poiseuille)}
\begin{enumerate}
    \item Measure $\langle v^y(\bm{r})\rangle$ in an atomic system.
    \item Fit the obtained $\langle v^y(\bm{r})\rangle$ to Eq.~(\ref{eq:pois_uptosecond_main}), using $\eta_0$ and $A_{\rm P}$ as fitting parameters.
\end{enumerate}
In this protocol, the mean density $\rho_0$, external force $g$, temperature $T$, and system size $L$ in Eq.~(\ref{eq:pois_uptosecond_main}) are fixed to the values used in the atomistic simulation.

\begin{figure}[tb]
\begin{center}
\includegraphics[width=\textwidth]{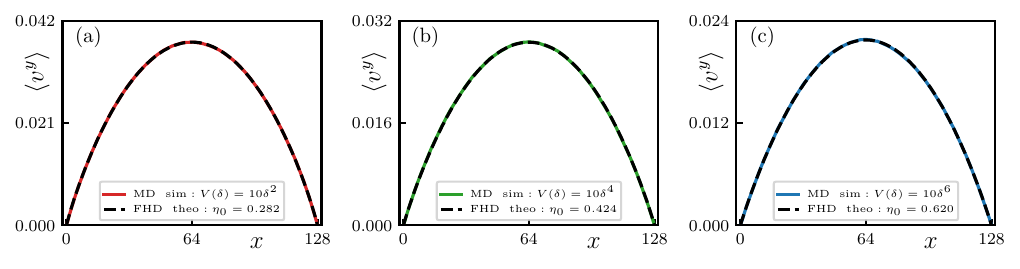}
\end{center}
\caption{
Estimation of the bare viscosity using Protocol 2 in the Poiseuille setup.
Results are shown for three atomic systems: (a) $V(\delta) = 10\delta^2$, (b) $V(\delta) = 10\delta^4$, and (c) $V(\delta) = 10\delta^6$.
The parameters are $\rho_0 = 0.765$, $T = 1.0$, and $L = 128.0$.
The colored curves show MD velocity profiles for an external force $g=0.00001$, and the black dashed curves show fits to the fluctuating-hydrodynamic expression in Eq.~(\ref{eq:pois_uptosecond_main}) over the entire system.
The estimated bare viscosities are summarized in Table~\ref{suptab3}.
}
\label{supfig8}
\end{figure}

We apply this protocol to the three atomic systems studied in Figs.~\ref{fig2}, \ref{supfig6} and \ref{supfig7}.
Figure~\ref{supfig8} shows the MD velocity profiles and the corresponding fits to Eq.~(\ref{eq:pois_uptosecond_main}).
The analytical expression captures the velocity profiles for all three systems.

The fitted bare viscosities are summarized in Table~\ref{suptab3}.
We perform the fits either over the entire velocity field or only near the wall.
Both procedures give values close to those obtained from Protocol 1 in the Couette geometry, although the Poiseuille estimates are systematically slightly smaller.

\begin{table}[b]
\centering
\begin{tabular*}{0.95\columnwidth}{@{\extracolsep{\fill}}c|c|c|c}
\hline
Atomic System & Couette & \multicolumn{2}{c}{Poiseuille} \\
 & (Protocol 1)  & (All Regions) & (Near Wall) \\
\hline
$V(\delta) = 10 \delta^2$ & 0.325 & 0.282 & 0.298 \\
$V(\delta) = 10 \delta^4$ & 0.470 & 0.424 & 0.449 \\
$V(\delta) = 10 \delta^6$ & 0.660 & 0.620 & 0.646 \\
\hline
\end{tabular*}
\caption{
Bare viscosities estimated from the Poiseuille setup for three atomic systems.
For comparison, the Couette values obtained from Protocol 1 in the main text are also shown.
The Poiseuille values are obtained by fitting the velocity field to Eq.~(\ref{eq:pois_uptosecond_main}).
``All Regions'' denotes fits over the entire system, whereas ``Near Wall'' denotes fits within 12 particle layers from the wall.
}
\label{suptab3}
\end{table}

\subsection{Derivation of the analytic expressions}

We derive Eq.~(\ref{eq:pois_uptosecond_main}).
We use the same simplified fluctuating-hydrodynamic equation as in Appendix~\ref{app4-1}, with the body force added:
\begin{align}
    \pd{\bm{v}}{t} + \epsilon \bm{v} \cdot \nabla \bm{v} &= \nu_0 \Delta \bm{v} + g \bm{e}_y + \bm{f}, \label{eq: iNSeqP, to solve} \\
    \langle f^a(\bm{r},t) f^b(\bm{r}',t')\rangle &= - \frac{2\nu_0 k_B T}{\rho_0} \delta^{ab} \Delta \delta(\bm{r}-\bm{r}') \delta(t-t'), \label{eq: irstP, to solve}
\end{align}
where $g$ represents the constant external force per unit mass, and $\bm{e}_y$ is the unit vector in the $y$ direction.
The no-slip boundary condition is
\begin{equation}
    \bm{v} = \bm{0} \quad {\rm at} \ \ x = 0, L.
    \label{eq: BCappP}
\end{equation}
Periodic boundary conditions are imposed along the $y$ axis.

We decompose the fluctuating velocity field $\bm{v}(\bm{r},t)$ into a noise-averaged component $\mybra \bm{v}(\bm{r},t) \myket$ and fluctuations $\bm{u}(\bm{r},t)$ around it:
\begin{align}
   v^a(\bm{r},t) &= \mybra v^a(\bm{r},t) \myket + u^a(\bm{r},t).
\end{align}
The boundary conditions given by Eq.~(\ref{eq: BCappP}) become:
\begin{align}
\mybra \bm{v} \myket &= \bm{0} \quad {\rm at} \ \ x= 0, L,
\label{eq: BCappP: det} \\
\bm{u} &= \bm{0} \quad {\rm at} \ \ x=0 , L .
\label{eq: BCappP: flu}
\end{align}
In the steady state considered here, the mean flow is unidirectional: $\langle v^x(\bm{r})\rangle=0$, and $\langle v^y(\bm{r})\rangle$ depends only on the wall-normal coordinate $x$.
Averaging the momentum balance as in Appendix~\ref{app4-2}, we obtain
\begin{align}
    \partial_b \left[\epsilon \Big(\langle v^a \rangle \langle v^b \rangle + \langle u^a u^b \rangle\Big) - \nu_0 \partial_b \langle v^a \rangle \right] = g \delta_{ay}.
\end{align}
The $y$ component gives
\begin{align}
     \nu_0 \partial_x \langle v^y \rangle = \epsilon \langle u^x u^y \rangle - g x + C,
    \label{eq: stressbalanceP}
\end{align}
where $C$ is an integration constant determined by the boundary conditions.
The velocity fluctuations $\bm{u}(\bm{r},t)$ obey
\begin{align}
    \frac{\partial \bm{u}}{\partial t}
    + \epsilon \Big(
    \mybra \bm{v} \myket \cdot \nabla \bm{u}
    + \bm{u} \cdot \nabla \mybra \bm{v} \myket
    + \bm{u} \cdot \nabla \bm{u}
    - \mybra \bm{u} \cdot \nabla \bm{u} \myket\Big)
    &= \nu_0 \Delta \bm{u} + \bm{f}.
    \label{eq: iNSeq_fluP, to solve}
\end{align}
Solving Eqs.~(\ref{eq: stressbalanceP}) and (\ref{eq: iNSeq_fluP, to solve}) with Eqs.~(\ref{eq: BCappP: det}) and (\ref{eq: BCappP: flu}) determines $\langle v^y \rangle$.

We now expand in $\epsilon$ [Eqs.~(\ref{eq:expand_det}) and (\ref{eq:expand_flu})].
Keeping terms through $O(\epsilon^2)$, the equations needed to compute $\langle v^y_{(\leq 2)}\rangle$ are
\begin{align}
     \nu_0 \partial_x \langle v^y \rangle
     &= C + \epsilon \langle u^x_{(0)} u^y_{(0)} \rangle
     - g x + \epsilon^2
     \Big[
     \langle u^x_{(1)} u^y_{(0)} \rangle
     + \langle u^x_{(0)} u^y_{(1)} \rangle
     \Big],
    \label{eq:pois_uptosecond}
\end{align}
and
\begin{align}
    \frac{\partial u^a_{(0)}}{\partial t} &= \nu_0 \Delta u^a_{(0)} + f^a,
    \label{eq:pois_flu_zero}
\end{align}
\begin{align}
    \frac{\partial u^a_{(1)}}{\partial t}
    &+ \Big[\langle v^y_{(0)}\rangle \partial_y u_{(0)}^a
    + u_{(0)}^b \partial_b \langle v^y_{(0)}\rangle + u_{(0)}^b \partial_b u_{(0)}^a
    - \langle u_{(0)}^b \partial_b u_{(0)}^a \rangle \Big]
    = \nu_0 \Delta u^a_{(1)}
    \label{eq:pois_flu_first}
\end{align}
with
\begin{align}
    \langle v^y_{(0)}\rangle = \frac{g}{2\nu_0}x(L-x).
\end{align}
The corresponding boundary conditions are
\begin{equation}
    \langle v^y_{(\leq 2)} \rangle = 0 \quad {\rm at} \ \ x=0, L
    \label{eq:pois_bou_second}
\end{equation}
\begin{equation}
        \bm{u}_{(0)} = \bm{0} \quad {\rm at} \ \ x=0,L,
        \qquad
        \bm{u}_{(1)} = \bm{0} \quad {\rm at} \ \ x=0,L.
    \label{eq:pois_flubou_first}
\end{equation}

To evaluate the fluctuation correlations, we use the Green function in Eq.~(\ref{eq:Green_expli}).
The zeroth- and first-order fluctuation fields are represented as
\begin{align}
    \uz^a(\bm{r},t)
    &= \int G(\bm{r},\bm{r}',\tau)f^a(\bm{r}',t-\tau),
    \label{eq:solzeroP}\\
    \uf^a(\bm{r},t)
    &= -\int G(\bm{r},\bm{r}',\tau) \biggl[
    \frac{g}{2\nu_0}x'(L-x')\partial_{y'} u_{(0)}^a
    + \frac{g}{2\nu_0}(L-2x') u_{(0)}^x \delta_{ay} \nonumber\\
    &\quad + u_{(0)}^b \partial_b u_{(0)}^a
    - \langle u_{(0)}^b \partial_b u_{(0)}^a \rangle
    \biggr]\biggr\vert_{\bm{r}',t-\tau},\label{eq:solfirstP}
\end{align}
where $\int=\int^\infty_0 d\tau\int^L_0 dx'\int^\infty_{-\infty}dy'$.
Since the zeroth-order field is Gaussian, we have
\begin{align}
    \langle  u_{(0)}^x u_{(0)}^y \rangle = 0,\\
    \langle  u_{(0)}^a u_{(0)}^b u_{(0)}^c \rangle = 0.
\end{align}
The first identity eliminates the $O(\epsilon)$ term in Eq.~(\ref{eq:pois_uptosecond}).
Using Eq.~(\ref{eq:solfirstP}) and the vanishing of the three-point function, we obtain
\begin{align}
    \langle \uf^a(\bm{r},0)\uz^b(\bm{r},0)\rangle
    &= - \frac{g}{2\nu_0}\int G(\bm{r},\bm{r}',\tau) \biggl[
    x'(L-x')\langle \uz^b(\bm{r},0)
    \partial_{y'} \uz^a(\bm{r}',-\tau)\rangle \nonumber \\
    &\quad + (L-2x') \delta_{ay}\langle \uz^b(\bm{r},0)
    \uz^x(\bm{r}',-\tau) \rangle
    \biggr],
\end{align}
and the components entering Eq.~(\ref{eq:pois_uptosecond}) are
\begin{align}
    \langle \uf^x(\bm{r},0)\uz^y(\bm{r},0)\rangle
    &= 0, \label{eq:u1xu0yP} \\
    \langle \uf^y(\bm{r},0)\uz^x(\bm{r},0)\rangle
    &= - \frac{g}{2\nu_0} \int (L-2x')
    G(\bm{r},\bm{r}',\tau)
    \langle \uz^x(\bm{r},0)
    \uz^x(\bm{r}',-\tau) \rangle \nonumber \\
    &= \frac{\sqrt{2} g k_B T}{4 \nu_0^2 L}
    \sum_{k_x, k'_x}
    \frac{a(k_x, k'_x)}{\sqrt{k_x^2 + (k'_x)^{2}}}
    \sin(k_x x) \sin(k'_x x)
    \label{eq:u0xu1yP}
\end{align}
with
\begin{align}
    a(k_x, k'_x) = \frac{2}{L} \int_0^L dx (2x - L) \sin(k_x x) \sin(k'_x x) = \frac{4}{L} \left[\frac{1}{(k_x + k'_x)^2} - \frac{1}{(k_x - k'_x)^2} \right]
\end{align}
for $n+n'={\rm odd}$, and zero otherwise.
Here $k_x=\pi n/L$ and $k'_x=\pi n'/L$.
Substituting this result into Eq.~(\ref{eq:pois_uptosecond}) gives
\begin{align}
     \nu_0 \partial_x \langle v^y_{(\leq 2)}(\bm{r})\rangle
     &= C - g x + \epsilon^2 \frac{\sqrt{2} g k_B T}{4\nu_0^2 L}
     \sum_{k_x, k'_x}
     \frac{a(k_x, k'_x)}{\sqrt{k_x^2 + (k'_x)^2}} \sin(k_x x) \sin(k'_x x).
\end{align}

Integrating with respect to $x$ and using the boundary conditions, we obtain
\begin{align}
    \langle v^y_{(\leq 2)}(\bm{r})\rangle
    &= \frac{g}{2\nu_0}x(L-x)
    + \epsilon^2 \frac{\sqrt{2} g k_B T}{8\nu_0^3 L}
    \sum_{k_x, k'_x}
    \frac{a(k_x, k'_x)}{\sqrt{k_x^2 + (k'_x)^2}} \nonumber \\
    &\qquad {}\left[
    \frac{\sin((k_x-k'_x)x)}{k_x-k'_x}
    - \frac{\sin((k_x+k'_x)x)}{k_x+k'_x}
    \right].
    \label{eq:pois_uptosecond_int}
\end{align}
Setting $\epsilon=1$ and using $\nu_0=\eta_0/\rho_0$, this expression reproduces Eq.~(\ref{eq:pois_uptosecond_main}) with the definition of $A_{\rm P}$.

\bibliographystyle{apsrev4-1}
\bibliography{
./bib/ref_basis.bib,
./bib/ref_anom_transp.bib,
./bib/ref_bare_param.bib,
./bib/ref_compu_fhd.bib,
./bib/ref_donev.bib,
./bib/ref_lrc.bib,
./bib/ref_noneq_ptrans.bib,
./bib/ref_other_app.bib,
./bib/ref_renor.bib,
./bib/ref_turbu.bib,
./bib/ref_2dexp.bib
}

\end{document}